\documentclass[journal=jctcce,manuscript=article]{achemso}

\usepackage[version=3]{mhchem}
\usepackage{color}
\usepackage{amssymb}

\author{Csaba F\'abri}
\email{ficsaba@staff.elte.hu}
\affiliation{HUN-REN--ELTE Complex Chemical Systems Research Group, P.O. Box 32, H-1518 Budapest 112, Hungary}
\alsoaffiliation{Department of Theoretical Physics, University of Debrecen, P.O. Box 400, H-4002 Debrecen, Hungary}

\author{G\'abor J. Hal\'asz}
\affiliation{Department of Information Technology, University of Debrecen, P.O. Box 400, H-4002 Debrecen, Hungary}

\author{Jaroslav Hofierka}
\affiliation{Theoretische Chemie, Physikalisch-Chemisches Institut, Universit\"at Heidelberg, D-69120 Heidelberg, Germany}

\author{Lorenz S. Cederbaum}
\affiliation{Theoretische Chemie, Physikalisch-Chemisches Institut, Universit\"at Heidelberg, D-69120 Heidelberg, Germany}

\author{\'Agnes Vib\'ok}
\email{vibok@phys.unideb.hu}
\affiliation{Department of Theoretical Physics, University of Debrecen, P.O. Box 400, H-4002 Debrecen, Hungary}
\alsoaffiliation{ELI-ALPS, ELI-HU Non-Profit Ltd, H-6720 Szeged, Dugonics t\'er 13, Hungary}

\title{Impact of dipole self-energy on cavity-induced nonadiabatic dynamics}

\begin{document}

\begin{abstract}
The coupling of matter to the quantized electromagnetic field of a plasmonic or 
optical cavity can be harnessed to modify and control chemical and physical 
properties of molecules. In optical cavities, a term known as the dipole 
self-energy (DSE) appears in the Hamiltonian to assure gauge invariance. 
The aim of this work is twofold. First, we introduce a method, 
which has its own merits and complements existing methods, to compute the DSE. Second, we 
study the impact of the DSE on cavity-induced nonadiabatic dynamics in a realistic system.
For that purpose, various matrix elements of the DSE are computed as functions of 
the nuclear coordinates and the dynamics of the system after laser excitation
is investigated. The cavity is known to induce conical intersections between
polaritons, which gives rise to substantial nonadiabatic effects. The DSE is
shown to slightly affect these light-induced conical intersections and, 
in particular, break their symmetry.
\end{abstract}

\section{Introduction}
\label{sec:intro}

Molecular polaritonics is an emerging and rapidly developing field
of research.\cite{15ToBa,17FlRuAp,18RiMaDu,18FeGaGa,18FlRiNa,19HeWaMo,19YuMe,20HeOw,21GaFrCi,22FrGaFe,22LiCuSu,23MaTaWe,23BhMoKo}
The quantized radiation field of optical or plasmonic nanocavities
can interact with the dipole moment of molecules,
which gives rise to new hybrid light-matter (so-called polaritonic) states,
carrying both excitonic (vibronic) and photonic features. 
Depending on their characteristic frequencies, quantized radiation modes can
couple to electronic or vibrational degrees of freedom, leading to
vibronic or vibrational polaritons \cite{20HeOw}.
Over the last decade, an array of experimental \cite{12HuScGe,16ChNiBe,16Ebbesen,16ThGeSh,16VeGeCh,16ZhChWa,17ChThAk,18GrTo,19DaWeKr,19OjChDe,19RoShEr,19VeThNa}
and theoretical 
\cite{15GaGaFe,16GaGaFe,16KoBeMu,17FlApRu,17HeSp_2,18FeGaGa,18PiScCh,18RuTaFl,18SzHaCs_2,18Vendrell,19MaHu,20MaMoHu,20TaMaZh,20HaRoKj,20LiNiSu,20LiSuNi,21LiMaHu,21AhHuBe,24Fabri,23ScKo_2,23ScKo,23ScSiRu,24FaCsHa,24FiRi,22RiHaRo,23BaDr,23Szidarovszky,23Szidarovszky_2,24AkVa}
works have demonstrated that polaritonic states can dramatically modify or even
control the physical and chemical properties of molecules. 
Indeed, the presence of the quantized radiation field can amplify \cite{24SaFeGa}
or suppress \cite{16GaGaFe} certain physical processes and even induce
new ones. We mention the possibility of strongly varying the rate
of spontaneous emission \cite{21Wang}, or influencing the rate
of energy transfer \cite{18DuMaRi,21LiNiSu} and charge transfer processes
\cite{19SeNi,20MaKrHu,21WePuSc}. In chemistry, the formation of molecular
polaritons offers the possibility of controlling chemical reactions \cite{16KoBeMu,19CaRiZh,19GaClGa,22LiCuSu,23FrCo}
such as photochemical reactivity \cite{20FeFrSc,20FrCoPe}, photoisomerization
\cite{18FrGrCo,20FrGrPe}, photodissociation \cite{20DaKo_2,21ToFe},
photoionization \cite{23FaHaCe} or photoassociation \cite{24CeFe}
of molecules. 

The strong coupling regime is reached when the rate of energy exchange
between cavity photons and the molecule is larger than the rate of
photon loss and system dephasing. Strong coupling can be achieved easier with 
plasmonic nanocavities since plasmonic modes are confined
to much smaller volumes than optical cavity photons. However, the
lifetime of plasmonic modes is much shorter due to high photon leakage.
Modes in plasmonic nanostructures almost always correspond to material 
excitations (see also Refs. \citenum{08ScBu,13Buhmann,21FeFeGa} 
for the quantization of medium-assisted electromagnetic fields).
Therefore, the interaction between plasmonic modes and molecules is
mostly guided by the Coulomb interaction which is hardly affected by the 
Power--Zienau--Woolley transformation.\cite{59PoZiMa,14VuGrDo,20Woolley}
Consequently, the cavity-molecule interaction 
can be described with sufficient
accuracy by using the usual $\vec{\mathrm{E}}\cdot\vec{\mathrm{d}}$ term 
(electric dipole approximation), without including the dipole self-energy (DSE).
In contrast, for electromagnetic fields in optical or Fabry--P\'erot
cavities, only the transverse component of the vector potential exists, therefore, 
the DSE term should be included in the interaction term.\cite{18RoWeRu}
Hereafter, we will consider the electromagnetic confinement to be an optical
cavity. Although strong coupling in optical cavities is currently 
feasible only for molecular ensembles in experiments, our computations
assume a single molecule coupled to an optical cavity in the strong coupling regime. 

Light-matter interaction can also induce nonadiabatic effects that are
similar in nature to natural nonadiabatic phenomena.\cite{17CsHaCe_2,18SzHaCs_2,19CsKoHa,20GuMu,20GuMu_2,20SzHaVi,20FaLaHa,20FaLaHa_2,21FaHaCe,21FaHaCe_2,21SzBaHa,21FaMaHu,22BaUmFa,22CsVeHa,22FaHaCe,22FaHaVi,22FiSa,24FaCsHa_2}
In case of light-induced nonadiabaticity, the radiation field strongly 
mixes the vibrational, rotational and electronic degrees of freedom of molecules. 
As a result, light-induced conical intersections (LICIs) are created between
polaritonic potential energy surfaces (PESs), leading to the breakdown of the
Born–Oppenheimer (BO) approximation.\cite{27BoOp}
In optical cavities, the DSE term  requires the evaluation of matrix elements 
of the squared dipole moment in the basis of electronic states, which is not
available in standard electronic structure program packages.
Current techniques for treating the DSE include the resolution-of-identity
approach\cite{24BoScKo} or approximating the DSE term with the square of the
ground-state molecular dipole moment for vibrational polaritons.\cite{24YuBo}
Here, we propose a method that we expect to be more accurate and robust for computing 
the DSE term. Namely, electronic matrix elements of the squared dipole moment are evaluated
directly in the basis of the ground and excited electronic states using the toolbox of 
electronic structure theory. Thus, no additional approximations are introduced in our 
treatment on top of the  standard ones used in quantum chemistry. To the best of our
knowledge, this is the first such implementation treating both the ground and excited
electronic states.
It is important to note that a similar approach was employed in Ref. \citenum{24FiRi} 
where the ground-state expectation value of the squared dipole moment was obtained 
directly using Hartree--Fock theory.
We also mention that the DSE term is taken into account in 
quantum-electrodynamic electronic-structure methods.\cite{20HaRoKj,21YaOuPe,23RuSiRu,23ScKo}

In this work, we demonstrate how the inclusion of the DSE term affects 
light-induced nonadiabatic quantum dynamics. Special emphasis is placed on
how the position of the LICI changes due to the DSE term and what quantum-dynamical 
consequences  (such as modified conditions for resonant  cavity-molecule coupling)
the DSE has for a single molecule coupled to an optical cavity mode. 
The structure of the paper is as follows. Section \ref{sec:theory} provides the theoretical description
of the cavity-molecule system and the details of electronic structure computations
necessary to  incorporate the DSE in our procedure.  In Section \ref{sec:compdet}
the computational models are described. Section \ref{sec:results} presents and discusses
the results of the current study, while conclusion and outlook are given in 
Section \ref{sec:summary}.

\section{Theory}
\label{sec:theory}

\subsection{Cavity-molecule system}
A single molecule interacting with a cavity mode is described by the
Pauli--Fierz Hamiltonian\cite{04CoDuGr}
\begin{equation}
        \hat{H} = \hat{H}_0 + \hbar \omega_\textrm{c} \hat{a}^\dag \hat{a} - g \hat{\vec{\mu}} \vec{e} (\hat{a}^\dag + \hat{a}) +
            \frac{g^2}{\hbar \omega_\textrm{c}} (\hat{\vec{\mu}} \vec{e})^2
   \label{eq:Hcm}
\end{equation}
where $\hat{H}_0$ is the Hamiltonian of the isolated (field-free) molecule, $\omega_\textrm{c}$ denotes the cavity-mode angular frequency, $\hat{a}^\dag$ and $\hat{a}$
are creation and annihilation operators of the cavity mode, $\hat{\vec{\mu}}$ is the
electric dipole moment operator of the molecule and $\vec{e}$ corresponds to the cavity
field polarization vector. The cavity-molecule coupling can be characterized by
the coupling strength parameter $g = \sqrt{\frac{\hbar \omega_\textrm{c}}{2 \epsilon_0 V}}$
where $\epsilon_0$ and $V$ are the permittivity and quantization volume of the cavity, 
respectively. The last term of $\hat{H}$ is the so-called dipole self-energy
(DSE).\cite{18RoWeRu,20ScRuRo,23SiScOb,23ScSiRu}

One way of treating cavity-molecule interactions is the cavity Born--Oppenheimer 
approach (CBOA).\cite{17FlApRu,23ScKo,23ScSiRu,23AnHaRo,24FiRi}
In the CBOA, photonic modes are treated nuclear-like and there is a formal equivalence 
between nuclear and photonic degrees of freedom. In contrast, photonic degrees of 
freedom can be also grouped with electrons (polaritonic approach) which is particularly 
useful when multiple electronic states are involved in the dynamics and nonadiabatic
effects have to be considered.\cite{23FiSa} We focus here on the latter.

If we consider two molecular electronic states (labeled by X and A), 
the matrix representation of $\hat{H}$ can be expressed as
\begin{equation}
    \hat{H}_\textrm{cm} =
    \begin{bmatrix}
            \hat{T} + V_\textrm{X} + Y_\textrm{X}& Y_{\textrm{XA}} & W_{\textrm{X}}^{(1)} & W_{\textrm{XA}}^{(1)} & \dots \\
            Y_{\textrm{XA}} & \hat{T} + V_\textrm{A} + Y_\textrm{A} & W_{\textrm{XA}}^{(1)} & W_{\textrm{A}}^{(1)} & \dots \\
            W_{\textrm{X}}^{(1)} & W_{\textrm{XA}}^{(1)} & \hat{T} + V_\textrm{X} + Y_\textrm{X} + \hbar\omega_\textrm{c} & Y_{\textrm{XA}} &\dots \\
            W_{\textrm{XA}}^{(1)} & W_{\textrm{A}}^{(1)} & Y_{\textrm{XA}} &\hat{T} + V_\textrm{A}  + Y_\textrm{A} + \hbar\omega_\textrm{c} & \dots \\
            \vdots & \vdots & \vdots & \vdots & \ddots 
        \end{bmatrix}
    \label{eq:cavity_H}
\end{equation}
in the direct product basis of electronic states $|\alpha\rangle$ ($\alpha = \textrm{X}, \textrm{A}$) and
Fock states of the cavity mode $|n\rangle$ ($n=0,1,2,\dots$). 
$\hat{H}_\textrm{cm}$ is an operator in the space of nuclear coordinates, 
$\hat{T}$ is the nuclear (rotational-vibrational) kinetic energy operator, 
while $V_\textrm{X}$ and $V_\textrm{A}$ are the ground-state 
and excited-state molecular potential energy surfaces (PESs).
The coupling between the cavity field and the molecular dipole moment operator
is described by the operators 
$W_\alpha^{(n)} = -g \sqrt{n} \langle \alpha  | \hat{\vec{\mu}} \vec{e} | \alpha \rangle$
($\alpha = \textrm{X}, \textrm{A}$) and
$W_{\textrm{XA}}^{(n)} = -g\sqrt{n}\langle\textrm{X}|\hat{\vec{\mu}}\vec{e}|\textrm{A}\rangle$.
Finally, the contribution of the DSE is given by the terms
\begin{equation}
    Y_\alpha = \frac{g^2}{\hbar \omega_\textrm{c}}\langle \alpha | (\hat{\vec{\mu}} \vec{e})^2 | \alpha \rangle
\label{eq:dse1}
\end{equation}
for $\alpha = \textrm{X}, \textrm{A}$ and
\begin{equation}
    Y_\textrm{XA} = \frac{g^2}{\hbar \omega_\textrm{c}}\langle \textrm{X} | (\hat{\vec{\mu}} \vec{e})^2 | \textrm{A} \rangle.
\label{eq:dse2}
\end{equation}
In Eq. \eqref{eq:cavity_H} $\hat{H}_\textrm{cm}$ is expressed in the so-called diabatic
representation (that is, cavity-molecule coupling terms appear in the potential energy).
Polaritonic (hybrid light-matter) PESs are defined as eigenvalues of the potential energy part of $\hat{H}_\textrm{cm}$ at each nuclear configuration.
In what follows, we will assume resonant coupling between electronic states X and A
(in other words, the cavity frequency is tuned in resonance with the 
$\textrm{X}\rightarrow\textrm{A}$ electronic transition).
We retain only the resonantly coupled X and A states and neglect other
electronic states which might be necessary to get converged results at higher
coupling strengths.\cite{21YaOuPe}

Effects associated with cavity loss are taken into account by the non-Hermitian time-dependent Schr\"odinger equation (TDSE)\cite{95ViNi}
\begin{equation}
\textrm{i} \hbar \frac{\partial |\Phi\rangle}{\partial t} =
    \Bigl(\hat{H}'-\textrm{i} \frac{\gamma_\textrm{c}}{2}\hat{N}\Bigr)|\Phi\rangle
	\label{eq:Schrodinger}
\end{equation}
where $\hat{N} = \hat{a}^\dag \hat{a}$ and
$\gamma_\textrm{c}$ specifies the cavity decay rate (equivalent to a finite 
photon lifetime of $\tau = \hbar/\gamma_\textrm{c}$).
It is worth noting that since we work in the length gauge (or length 
representation), photonic and material degrees of freedom are mixed. As a consequence,
self-polarization enters the definition of the photon number operator and 
$\hat{a}^\dag \hat{a}$ is not the physical photon number.\cite{18RoWeRu,20ScRuRo}
In Eq. \eqref{eq:Schrodinger}, $\hat{H}'$ includes a term which describes the 
interaction of the molecule with  a laser pulse, that is, 
$\hat{H}' = \hat{H} - \hat{\vec{\mu}} \vec{E}(t)$.
The laser pulse is specified by $|\vec{E}(t)| = E_0 \sin^2(\pi t / T) \cos(\omega t)$
for $0 \le t \le T$ and $\vec{E}(t) = \vec{0}$ otherwise. $E_0$, $T$ and $\omega$ denote
the amplitude, length and carrier angular frequency of the pump pulse, respectively.
Dissipation during the time evolution of an open quantum system is typically 
taken into account by using the Lindblad master equation.\cite{20Manzano}
However, it is shown in Ref. \citenum{95ViNi} that dissipation can also be 
described by the non-Hermitian TDSE which provides a viable alternative 
for the Lindblad master equation. An important merit of the non-Hermitian TDSE formalism 
is that  time evolution can be expressed in terms of the state vector $|\Psi\rangle$ 
instead of the density operator (Lindblad master equation), which greatly reduces the 
computational cost. In case of the non-Hermitian TDSE, the wave function norm decreases 
over time due to the term $-\textrm{i} \frac{\gamma_\textrm{c}}{2} \hat{N}$ in 
Eq. \eqref{eq:Schrodinger}. Therefore, in the current work, we choose to set the norm of
the wave function to one at each time propagation step, which is also applied by the
stochastic Schr\"odinger method.\cite{93MoCaDa} In Ref. \citenum{24FaCsHa} we found that
renormalization of the wave function when propagating the non-Hermitian TDSE tends to
give results that agree well with data obtained with the Lindblad master equation.

\subsection{Electronic structure computations for the dipole self-energy}

The projection of $\hat{\vec{\mu}}$ on the polarization vector can be expressed as
\begin{equation}
    \hat{\vec{\mu}} \vec{e} = \sum_{k=x,y,z} \hat{\mu}_k e_k
\label{eq:pol1}
\end{equation}
where $e_k$ are components of the normalized polarization vector
$\mathbf{e}=(e_x,e_y,e_z)$ with $|\mathbf{e}|^2=e_x^2+e_y^2+e_z^2=1$. 
With this, the squared dipole projection in the DSE term has the form
\begin{equation}
    (\hat{\vec{\mu}} \vec{e})^2 = \sum_{k,l=x,y,z} \hat{\mu}_k \hat{\mu}_l e_k e_l.
\label{eq:pol2}
\end{equation}
Eqs. \eqref{eq:pol1} and \eqref{eq:pol2} are general, that is, they describe 
arbitrary polarization.

As already stated in Section \ref{sec:intro}, we evaluate the matrix elements of
$(\hat{\vec{\mu}} \vec{e})^2$ in the basis of electronic states $| \alpha \rangle$ directly.
Our approach provides an alternative to existing techniques which are briefly 
summarized now. One way of computing the DSE is to use the approximation
\begin{equation}
    \langle \textrm{X} | \hat{\mu}_k \hat{\mu}_l | \textrm{X} \rangle \approx 
    \langle \textrm{X} | \hat{\mu}_k | \textrm{X} \rangle \langle \textrm{X} | \hat{\mu}_l | \textrm{X} \rangle
\end{equation}
which has been done for the electronic ground state X.\cite{24YuBo}
Another possibility is to insert an approximate resolution of identity in between 
$\hat{\mu}_k$ and $\hat{\mu}_l$,
\begin{equation}
    \langle \alpha | \hat{\mu}_k \hat{\mu}_l | \alpha' \rangle \approx 
    \sum_{\alpha''} \langle \alpha | \hat{\mu}_k | \alpha'' \rangle \langle \alpha'' |\hat{\mu}_l | \alpha' \rangle
\end{equation}
where $|\alpha''\rangle$ denotes electronic states included in the resolution of 
identity.\cite{24BoScKo}
This approach might necessitate the computation of many electronic states.
This is to be contrasted with the current method, which requires only the electronic 
states of interest. In what follows, we present both general considerations and a 
simplified approach which has been used to compute the DSE term in the current work.

\subsubsection{General approach}

In order to incorporate the DSE term in the cavity-molecule Hamiltonian, one has to 
evaluate matrix elements of the operators $\hat{\mu}_k \hat{\mu}_l$ in the basis 
of electronic states (see also Eqs. \eqref{eq:dse1} and \eqref{eq:dse2}).
To the best of our knowledge, such matrix elements are not available in standard electronic
structure packages. Therefore, we have developed the necessary theory to obtain the desired
electronic matrix elements, which is described in this subsection.

The molecular dipole moment operator can be written as the sum of electronic 
($\hat{\mu}_k^\textrm{e}$) and nuclear ($\hat{\mu}_k^\textrm{n}$) contributions, that is,
\begin{equation}
    \hat{\mu}_k = \hat{\mu}_k^\textrm{e}+\hat{\mu}_k^\textrm{n} = 
                -\sum_{i=1}^{N_\textrm{e}} x_{ik} + \sum_{a=1}^{N_\textrm{n}} Z_a X_{ak}
\label{eq:muop}
\end{equation}
where $x_{ik}$ and $X_{ak}$ denote components of electronic and nuclear position vectors, 
and $Z_a$ is the charge of nucleus $a$. In addition, $N_\textrm{e}$ and $N_\textrm{n}$
refer to the number of electrons and nuclei in the (electrically neutral) molecule and
atomic units are assumed. Thus, we get
\begin{equation}
    \hat{\mu}_k \hat{\mu}_l = \hat{\mu}_k^\textrm{e} \hat{\mu}_l^\textrm{e}+
        \hat{\mu}_k^\textrm{e} \hat{\mu}_l^\textrm{n}+
        \hat{\mu}_l^\textrm{e}\hat{\mu}_k^\textrm{n}+
        \hat{\mu}_k^\textrm{n}\hat{\mu}_l^\textrm{n}
\label{eq:mu2op}
\end{equation}
where
\begin{equation}
    \hat{\mu}_k^\textrm{e} \hat{\mu}_l^\textrm{e} = 
        \sum_{i=1}^{N_\textrm{e}} \sum_{j=1}^{N_\textrm{e}} x_{ik} x_{jl}=
         \sum_{i=1}^{N_\textrm{e}} x_{ik} x_{il}+
         \sum_{i=1}^{N_\textrm{e}} \sum_{j \ne i}^{N_\textrm{e}} x_{ik} x_{jl}
\end{equation}
which includes one-electron and two-electron terms.
Using the indistinguishability of electrons, electronic matrix elements of 
$\mu_k^\textrm{e}\mu_l^\textrm{e}$ can be expressed as
\begin{equation}
    \langle \alpha | \hat{\mu}_k^\textrm{e} \hat{\mu}_l^\textrm{e} | \alpha' \rangle = 
    N_\textrm{e} \langle \alpha | x_{1k} x_{1l} | \alpha' \rangle +
    N_\textrm{e}(N_\textrm{e}-1) \langle \alpha | x_{1k}x_{2l} | \alpha' \rangle
  \label{eq:mue2}
\end{equation}
where $|\alpha\rangle$ and $|\alpha'\rangle$ denote molecular electronic states. 
It is obvious that the first and second terms of Eq. \eqref{eq:mue2} require the 
one-electron and two-electron reduced density ($|\alpha\rangle=|\alpha'\rangle$) or 
transition density ($|\alpha\rangle \ne |\alpha'\rangle$) matrices, respectively.

Similarly, matrix elements of $\hat{\mu}_k^\textrm{e}$ can be cast as
\begin{equation}
    \langle \alpha | \hat{\mu}_k^\textrm{e} | \alpha' \rangle = 
    -N_\textrm{e} \langle \alpha | x_{1k} | \alpha' \rangle
  \label{eq:mue}
\end{equation}
where only the one-electron reduced (transition) density matrix is needed.
With Eqs. \eqref{eq:mue2} and \eqref{eq:mue}, we get
\begin{equation}
    \langle \alpha | \hat{\mu}_k \hat{\mu}_l | \alpha' \rangle= 
    \langle \alpha | \hat{\mu}_k^\textrm{e} \hat{\mu}_l^\textrm{e} | \alpha' \rangle+
    \langle \alpha | \hat{\mu}_k^\textrm{e} | \alpha' \rangle \hat{\mu}_l^\textrm{n} +
    \langle \alpha | \hat{\mu}_l^\textrm{e} | \alpha' \rangle \hat{\mu}_k^\textrm{n} +
    \hat{\mu}_k^\textrm{n} \hat{\mu}_l^\textrm{n} \delta_{\alpha\alpha'}
  \label{eq:mu2}
\end{equation}
which can be evaluated employing any electronic structure method which provides the 
one-electron and two-electron reduced (transition) density matrices.

Another way of obtaining the expectation value of $\hat{\mu}_k \hat{\mu}_l$ is to consider the Hamiltonian
\begin{equation}
    \hat{H}(\lambda) = \hat{H}_0 + \lambda \hat{\mu}_k \hat{\mu}_l
  \label{eq:pt1}
\end{equation}
where the molecular Hamiltonian $\hat{H}_0$ is perturbed by the operator
$\hat{\mu}_k \hat{\mu}_l$ ($\lambda$ is the perturbation parameter).
Within the framework of this perturbation-based formalism it can be shown that
the expectation value of $\hat{\mu}_k \hat{\mu}_l$ over the unperturbed electronic wave function $| \alpha(0) \rangle$ equals
\begin{equation}
    \langle \alpha(0) | \hat{\mu}_k \hat{\mu}_l | \alpha(0) \rangle =
    \frac{\textrm{d}E(\lambda)}{\textrm{d}\lambda} \bigg|_{\lambda=0}
  \label{eq:pt2}
\end{equation}
where $E(\lambda)$ is the perturbation-dependent energy satisfying
$\hat{H}(\lambda) | \alpha(\lambda) \rangle = E(\lambda) | \alpha(\lambda) \rangle$.
Eq. \eqref{eq:pt2} is also known as the Hellmann--Feynman theorem\cite{17Jensen}
which can be extended to account for degenerate states.\cite{03AlCe} 
The Hellmann--Feynman approach is particularly useful if not all wave function 
parameters are optimized by the variational principle.\cite{17Jensen}

\subsubsection{Simplified approach}

In the current work, the excited state (A) is obtained by the configuration
interaction singles (CIS) method.\cite{92FoHePo} 
Using spin adaptation, singlet and triplet excited states can be computed separately by 
diagonalizing the singlet and triplet blocks of the CIS Hamiltonian matrix.\cite{92FoHePo}
The singlet block of the CIS Hamiltonian reads
\begin{equation}
    H_{ia,jb} = (\epsilon_a-\epsilon_i) \delta_{ij} \delta_{ab} + 
                2\langle ij | ab \rangle - \langle ia | jb \rangle
  \label{eq:cis}
\end{equation} 
where $i,j$ and $a,b$ label occupied and virtual spatial molecular 
orbitals (MOs), respectively.
Moreover, $\epsilon_a$ and $\epsilon_i$ are MO energies, while $\langle ij | ab \rangle$
and $\langle ia | jb \rangle$ refer to two-electron  (Coulomb) integrals in the basis 
of canonical MOs using physicist's notation (1212).
This way, CIS amplitudes $c_{ia}$ and excitation energies $\Delta E_\textrm{CIS}$
can be found by solving the eigenvalue equation
\begin{equation}
    \sum_{j}^\textrm{occ} \sum_{b}^\textrm{virt} H_{ia,jb} c_{jb} = \Delta E_\textrm{CIS} c_{ia}.
  \label{eq:cis2}
\end{equation}
The CIS energies of excited states are given by 
$E_\textrm{CIS} = E_\textrm{RHF}+\Delta E_\textrm{CIS}$ where $E_\textrm{RHF}$ is
the ground-state restricted Hartree--Fock (RHF) energy, 
while the singlet CIS wave functions assume the form
\begin{equation}
    | \Psi_\textrm{CIS} \rangle = 
        \frac{1}{\sqrt{2}} \sum_{i}^\textrm{occ} \sum_{a}^\textrm{virt} c_{ia}
        \left( |\Phi_{i\uparrow}^{a\uparrow}\rangle + |\Phi_{i\downarrow}^{a\downarrow} \rangle \right)
  \label{eq:cis_wf}
\end{equation}
where $|\Phi_{i\sigma}^{a\sigma}\rangle$ denote singly-excited Slater determinants with 
an electron of spin $\sigma=\uparrow,\downarrow$ promoted to the virtual MO $|a\rangle$ 
from the occupied MO $|i\rangle$. We emphasize that, due to Brillouin's theorem, 
singly-excited determinants do not mix with the ground-state RHF determinant.\cite{17Jensen}
As a consequence, the ground state at the CIS level of theory equals the RHF reference.

The CIS wave function is parameterized by the CIS amplitudes and MO coefficients,
out of which only the CIS amplitudes are determined variationally.
Therefore, the expectation value of $\hat{\mu}_k \hat{\mu}_l$ over $| \Psi_\textrm{CIS} \rangle$
will not equal the value obtained with the Hellmann--Feynman theorem.
In such cases, the Hellmann--Feynman approach is preferred to taking the
expectation value of the perturbing operator over the approximate wave function.
Thus, the excited-state property for the operator $\hat{\mu}_k \hat{\mu}_l$ is
obtained by the Hellmann--Feynman method introduced in Eq. \eqref{eq:pt2}.
The first step is to rewrite the CIS energy as
\begin{equation}
    E_\textrm{CIS} = E_\textrm{RHF} + \Delta E_\textrm{CIS} = E_\textrm{RHF} +
        \sum_{ij}^\textrm{occ} \sum_{ab}^\textrm{virt} H_{ia,jb} c_{ia} c_{jb}
\end{equation}
using Eq. \eqref{eq:cis2}. 
According to Eq. \eqref{eq:pt2}, we need to evaluate the derivative of the CIS energy
\begin{equation}
    \frac{\textrm{d}E_\textrm{CIS}(\lambda)}{\textrm{d}\lambda}\bigg|_{\lambda=0} = 
        \frac{\textrm{d}E_\textrm{RHF}(\lambda)}{\textrm{d}\lambda}\bigg|_{\lambda=0} +
        \frac{\textrm{d}\Delta E_\textrm{CIS}(\lambda)}{\textrm{d}\lambda}\bigg|_{\lambda=0}
\end{equation}
with respect to the perturbation parameter $\lambda$ at $\lambda=0$.
In order to be consistent with the notation $|\Phi_{i\sigma}^{a\sigma}\rangle$ used
for singly-excited Slater determinants, we denote the ground-state RHF wave function
as $| \Phi_0 \rangle$.
Since for the RHF method all wave function parameters (MO coefficients) are optimized by 
the variational principle, the derivative of the RHF energy is equal to the expectation 
value of $\hat{\mu}_k \hat{\mu}_l$ over $| \Phi_0 \rangle$, that is,
\begin{equation}
    \frac{\textrm{d}E_\textrm{RHF}(\lambda)}{\textrm{d}\lambda}\bigg|_{\lambda=0} =
        \langle \Phi_0 | \hat{\mu}_k \hat{\mu}_l | \Phi_0 \rangle
\end{equation}
which can be evaluated using Eq. \eqref{eq:mu2}. 
The derivative of the CIS excitation energy reads
\begin{equation}
    \frac{\textrm{d}\Delta E_\textrm{CIS}(\lambda)}{\textrm{d}\lambda}\bigg|_{\lambda=0} =
        \sum_{ij}^\textrm{occ} \sum_{ab}^\textrm{virt}
        \frac{\textrm{d}H_{ia,jb}(\lambda)}{\textrm{d}\lambda}\bigg|_{\lambda=0} c_{ia} c_{jb}.
    \label{eq:ddcis}
\end{equation}
Since the CIS amplitudes $c_{ia}$ diagonalize the CIS Hamiltonian, their derivatives do
not appear in Eq. \eqref{eq:ddcis}. According to Eq. \eqref{eq:cis}, matrix elements
$H_{ia,jb}$ depend on the MO energies and MOs. Therefore, one has to take the derivative 
of the MO energies and MOs with respect to $\lambda$, which has been accomplished by 
solving the coupled-perturbed Hartree--Fock (CPHF) equations.\cite{68GeMi}
A more efficient method to evaluate CIS energy derivatives is based on the
Lagrangian formalism for non-variational wave function parameters.\cite{17Jensen}

As the excited state is obtained at the CIS level of theory, we choose to work with the 
RHF ground state $| \Phi_0 \rangle$ for the sake of consistency. Therefore, transition
matrix elements of $\hat{\mu}_k \hat{\mu}_l$ are to be computed between 
$| \Phi_0 \rangle$ and the lowest singlet CIS excited state $|\Psi_\textrm{CIS} \rangle$.
This can be readily achieved by using the CIS expansion of Eq. \eqref{eq:cis_wf}, that is,
\begin{equation}
    \langle \Phi_0 | \hat{\mu}_k \hat{\mu}_l | \Psi_\textrm{CIS} \rangle = 
    \frac{1}{\sqrt{2}} \sum_{i}^\textrm{occ} \sum_{a}^\textrm{virt} c_{ia}
        \left(\langle \Phi_0 | \hat{\mu}_k \hat{\mu}_l |\Phi_{i\uparrow}^{a\uparrow}\rangle + 
        \langle \Phi_0 | \hat{\mu}_k \hat{\mu}_l |\Phi_{i\downarrow}^{a\downarrow} \rangle \right)
    \label{eq:cistr}
\end{equation}
where matrix elements between the RHF reference and singly-excited Slater 
determinants can be evaluated by the Slater--Condon rules.\cite{17Jensen}

\section{Computational details}
\label{sec:compdet}

\begin{figure}
\includegraphics[width=0.8\textwidth]{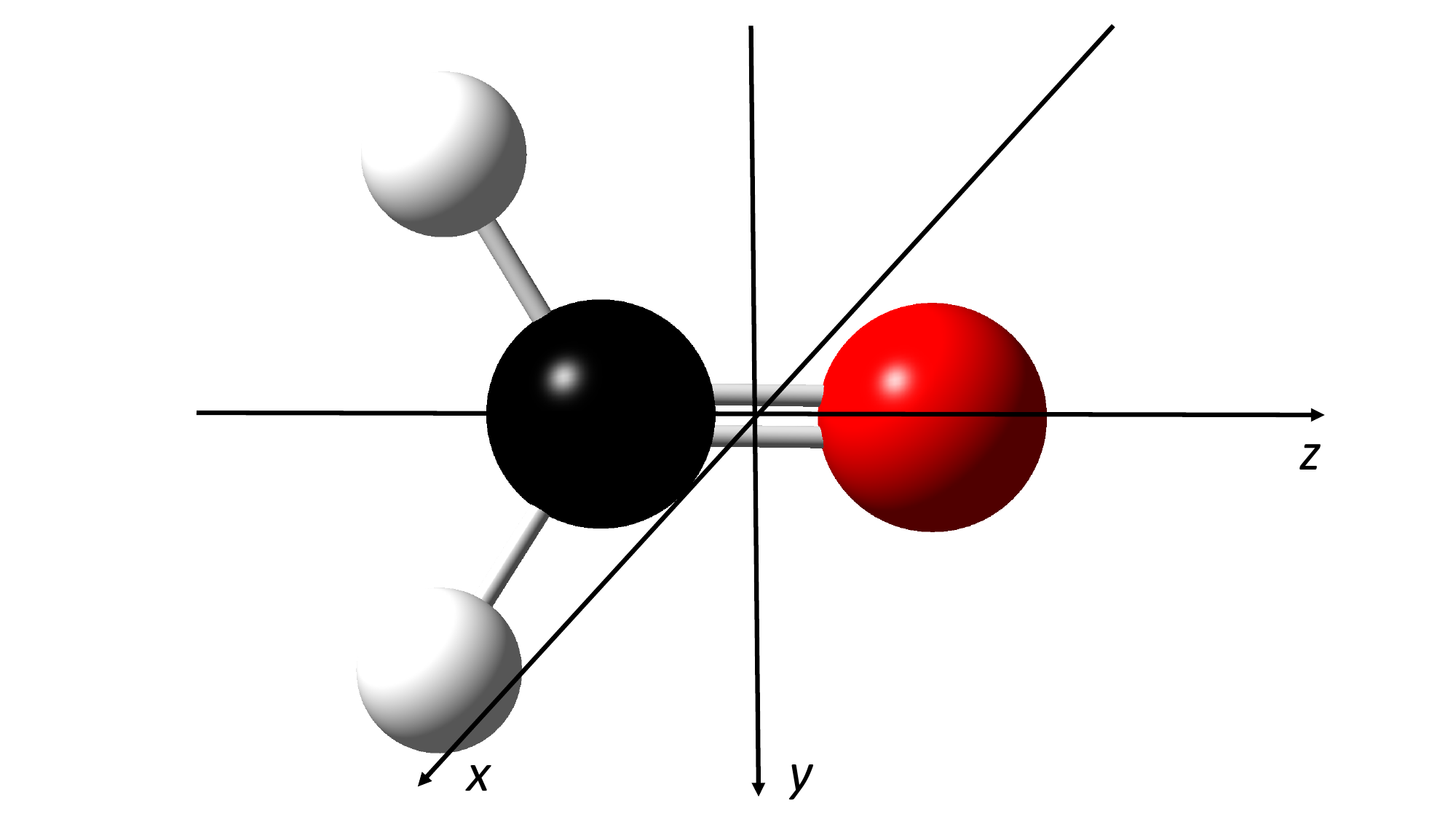}
\caption{\label{fig:structure} Reference structure of formaldehyde (H$_2$CO, 
$C_\textrm{2v}$ point group symmetry) and definition of body-fixed Cartesian axes.
The planar reference structure is placed in the $yz$ plane.}
\end{figure}

First, a brief description of the computational model is provided for the four-atomic formaldehyde (H$_{2}$CO) molecule.
Two singlet electronic states ($\textrm{S}_0 ~ (\tilde{\textrm{X}} ~ ^1\textrm{A}_1)$
and $\textrm{S}_1 ~ (\tilde{\textrm{A}} ~ ^1\textrm{A}_2)$) and two vibrational modes
($\nu_2$ (C=O stretch) and $\nu_4$ (out-of-plane)) are considered (2D($\nu_2$,$\nu_4$)
vibrational model,
see Refs. \citenum{20FaLaHa,20FaLaHa_2,21FaHaCe,21FaHaCe_2,22FaHaVi,22FaHaCe}
for further details).
In addition, the orientation of the molecule is fixed with respect to the cavity polarization vector $\vec{e}$. The reference structure of H$_2$CO ($C_\textrm{2v}$ point 
group symmetry) is shown in Fig. \ref{fig:structure} where the definition of body-fixed 
Cartesian axes is also given. For the 2D($\nu_2$,$\nu_4$) model, electronic structure 
computations were carried out at the DFT/TDDFT  (CAM-B3LYP/6-31G*) level of theory on a
two-dimensional grid along the $\nu_2$ and $\nu_4$ vibrational modes, yielding ground-state
(X) and excited-state (A) PESs, as well as permanent (PDM) and transition (TDM) dipole
moment surfaces.

The electronic structure data are now supplemented by matrix elements of the operators
$\hat{\mu}_k \hat{\mu}_l$ in the basis of electronic states. As these quantities are
not available in standard electronic structure packages, we propose an easy-to-implement
yet sensible approach based on wave-function methods.
The ground-state expectation value of $\hat{\mu}_k \hat{\mu}_l$ is evaluated using
Eqs. \eqref{eq:mue2}, \eqref{eq:mue} and \eqref{eq:mu2} at the RHF 
(restricted Hartree--Fock) level of theory. The excited state considered by the current 
work is the lowest singlet excited state of H$_2$CO, corresponding to a HOMO-LUMO excitation.
Thus, we have to deal with an open-shell singlet state which cannot be described by a 
single Slater determinant. Therefore, the configuration interaction singles (CIS) method 
is invoked to obtain the excited state. The excited-state property for
$\hat{\mu}_k \hat{\mu}_l$ and the matrix element 
$\langle \textrm{X} | \hat{\mu}_k \hat{\mu}_l | \textrm{A} \rangle$ are evaluated by 
Eqs. \eqref{eq:pt2} and \eqref{eq:cistr}, respectively. In all computations, 
we have used the 6-31G* basis set. The necessary integrals, canonical molecular orbitals,
orbital energies and RHF reduced density matrices have been computed with the PySCF
ab initio program package.\cite{18SuBeBl,20SuZhBa}
The methodology based on the RHF and CIS levels of theory has been validated by comparing 
several RHF/6-31G* and CIS/6-31G* quantities (excitation energies, PDM and TDM values) to 
their DFT and TDDFT (CAM-B3LYP/6-31G*) counterparts. The RHF/CIS and DFT/TDDFT methods have 
been found to give good agreement for all quantities considered.
 
At the reference structure, the X and A electronic states transform according to the
irreducible representations (irreps) $\Gamma_\textrm{X} = A_1$ and 
$\Gamma_\textrm{A} = A_2$ of the $C_\textrm{2v}$ point group, respectively, 
while the two vibrational modes are of $\Gamma_{\nu_2} = A_1$ and
$\Gamma_{\nu_4} = B_1$ symmetries. In addition, the body-fixed Cartesian coordinates 
transform according to the irreps $\Gamma_x = B_1$, $\Gamma_y = B_2$ and $\Gamma_z = A_1$.
Table \ref{tbl:grouptheory} provides a detailed symmetry analysis for electronic matrix
elements of the operators $\hat{\mu}_k$ and $\hat{\mu}_k \hat{\mu}_l$.
These matrix elements can have nonvanishing values at the reference geometry only in case 
of the $A_1$ irrep. Moreover, for the $B_1$ irrep, although matrix elements of 
$\hat{\mu}_k$ and $\hat{\mu}_k \hat{\mu}_l$ are zero at the reference geometry,
one can produce nonzero matrix elements by displacement from the reference structure
along the $\nu_4$ vibrational mode of $B_1$ symmetry.

\begin{table}
  \caption{Symmetry properties (irreducible representations of the $C_{2\textrm{v}}$ 
  point group) of electronic matrix elements 
  ($\langle \alpha |\hat{O}| \alpha \rangle$ with $\alpha = \textrm{X}, \textrm{A}$
  and $\langle \textrm{X} |\hat{O}| \textrm{A} \rangle$)
  for the operators $\hat{O}=\hat{\mu}_k$ and $\hat{O}=\hat{\mu}_k \hat{\mu}_l$
  ($k, l = x,y,z$). See Eqs. \eqref{eq:muop} and \eqref{eq:mu2op} for the 
  definition of $\hat{\mu}_k$ and $\hat{\mu}_k \hat{\mu}_l$.
  Note that $\Gamma_\textrm{X} = A_1$, $\Gamma_\textrm{A} = A_2$, $\Gamma_x = B_1$, 
  $\Gamma_y = B_2$ and $\Gamma_z = A_1$.}
  \label{tbl:grouptheory}
  \begin{tabular}{ccc}
  \hline
        operator ($\hat{O}$) & symmetry for $\langle \alpha |\hat{O}| \alpha \rangle$
        ($\alpha$ = X, A) & symmetry for $\langle \textrm{X} |\hat{O}| \textrm{A} \rangle$ \\
  \hline
        $\hat{\mu}_x$  & $\Gamma_\alpha \otimes \Gamma_x \otimes \Gamma_\alpha = B_1$ & $\Gamma_\textrm{X} \otimes \Gamma_x \otimes \Gamma_\textrm{A} = B_2$ \\
        $\hat{\mu}_y$  & $\Gamma_\alpha \otimes \Gamma_y \otimes \Gamma_\alpha = B_2$ & $\Gamma_\textrm{X} \otimes \Gamma_y \otimes \Gamma_\textrm{A} = B_1$ \\
        $\hat{\mu}_z$  & $\Gamma_\alpha \otimes \Gamma_z \otimes \Gamma_\alpha = A_1$ & $\Gamma_\textrm{X} \otimes \Gamma_z \otimes \Gamma_\textrm{A} = A_2$ \\
        $\hat{\mu}_x^2$  & $\Gamma_\alpha \otimes \Gamma_x \otimes \Gamma_x \otimes \Gamma_\alpha = A_1$ & $\Gamma_\textrm{X} \otimes \Gamma_x \otimes \Gamma_x \otimes \Gamma_\textrm{A} = A_2$ \\
        $\hat{\mu}_y^2$  & $\Gamma_\alpha \otimes \Gamma_y \otimes \Gamma_y \otimes \Gamma_\alpha = A_1$ & $\Gamma_\textrm{X} \otimes \Gamma_y \otimes \Gamma_y \otimes \Gamma_\textrm{A} = A_2$ \\
        $\hat{\mu}_z^2$  & $\Gamma_\alpha \otimes \Gamma_z \otimes \Gamma_z \otimes \Gamma_\alpha = A_1$ & $\Gamma_\textrm{X} \otimes \Gamma_z \otimes \Gamma_z \otimes \Gamma_\textrm{A} = A_2$ \\
        $\hat{\mu}_x \hat{\mu}_y$  & $\Gamma_\alpha \otimes \Gamma_x \otimes \Gamma_y \otimes \Gamma_\alpha = A_2$ & $\Gamma_\textrm{X} \otimes \Gamma_x \otimes \Gamma_y \otimes \Gamma_\textrm{A} = A_1$ \\
        $\hat{\mu}_x \hat{\mu}_z$  & $\Gamma_\alpha \otimes \Gamma_x \otimes \Gamma_z \otimes \Gamma_\alpha = B_1$ & $\Gamma_\textrm{X} \otimes \Gamma_x \otimes \Gamma_z \otimes \Gamma_\textrm{A} = B_2$ \\
        $\hat{\mu}_y \hat{\mu}_z$  & $\Gamma_\alpha \otimes \Gamma_y \otimes \Gamma_z \otimes \Gamma_\alpha = B_2$ & $\Gamma_\textrm{X} \otimes \Gamma_y \otimes \Gamma_z \otimes \Gamma_\textrm{A} = B_1$ \\
   \hline
\end{tabular}
\end{table}

In all quantum-dynamical computations, the wave function of the coupled cavity-molecule
system has been represented in the basis $|\alpha i n\rangle$ where 
$\alpha=\textrm{X},\textrm{A}$ labels molecular electronic states, $i$ denotes 
two-dimensional direct-product Fourier discrete variable representation 
(DVR)\cite{70Meyer} basis functions (used for the $\nu_2$ and $\nu_4$ vibrational modes) 
and $n = 0,1,2,3$ ($n$ labels Fock states of the cavity mode). 
Normal coordinates ($Q_2$ and $Q_4$) are used and the kinetic energy operator is equal to
$\hat{T} = -\frac{\hbar^2}{2} \left( \frac{\partial^2}{\partial Q_2^2}+
\frac{\partial^2}{\partial Q_4^2} \right)$.
$37$ and $51$ equidistant Fourier DVR points have been applied in the
intervals $[-110\sqrt{m_\textrm{e}} a_0,70\sqrt{m_\textrm{e}} a_0]$ and 
$[-125\sqrt{m_\textrm{e}} a_0,125\sqrt{m_\textrm{e}} a_0]$
for $Q_2$ and $Q_4$, respectively. The dimension of the two-dimensional direct-product DVR
basis equals $37 \cdot 51 = 1887$.
Time propagation of the wave function has been carried out with the explicit third-order 
Adams--Bashforth method\cite{03SuMa} with a time step of $0.005~\textrm{fs}$
and the norm of the wave function has been set to one
at each step of the time propagation.
As discussed in Section \ref{sec:theory}, the molecule is coupled to a single quantized 
mode of an optical cavity.
Then, the molecule is pumped with a laser pulse which couples the electronic states X and A.
Since in the  2D($\nu_2$,$\nu_4)$ model the X and A PDMs are perpendicular to the TDM and
the  polarization vector of the laser field is chosen parallel to the TDM, the pump pulse
interacts only with the TDM of the molecule.

\section{Results and discussion}
\label{sec:results}

Having described the theoretical details and computational models, we will now examine
electronic structure and quantum-dynamical results.
We consider single-molecule strong coupling conditions which imply that the ground 
(lowest) polaritonic state essentially equals $|\textrm{X}0\rangle$
(electronic ground state with zero photons), while the lower (1LP) and upper (1UP) 
polaritonic states can be well approximated as the
linear combinations of $|\textrm{X}1\rangle$ and $|\textrm{A}0\rangle$ 
(singly-excited subspace). 
Similarly, higher-lying polaritonic states can be denoted as $n$LP and $n$UP (with $n>1$). 
As already outlined in Section \ref{sec:compdet},
our computations are numerically exact including all cavity-molecule coupling terms
and the nomenclature is only used to guide the discussion.

After summarizing the electronic structure results relevant for the DSE term, 
we will discuss the impact of DSE on the light-induced conical intersection (LICI)
and the ensuing cavity-induced nonadiabatic effects. 
Thereby, time-dependent quantum-dynamical results with and without the
inclusion of the DSE term in the cavity-molecule Hamiltonian will be presented and
compared. To this end, the system is prepared in an initial state
which corresponds to the ground state of the coupled cavity-molecule system. 
Then, the system is excited by a pump laser pulse which transfers population primarily 
to the 1LP and 1UP polaritonic states. Time-dependent populations of the 1LP and 1UP 
states will highlight the impact of the DSE term.

\subsection{Electronic structure results}

\begin{figure}
\includegraphics[width=0.75\textwidth]{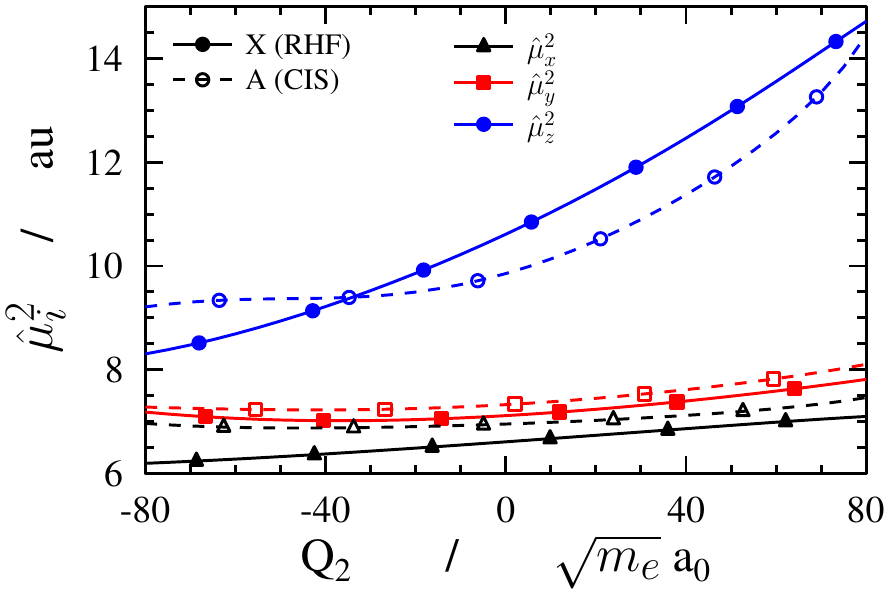}
\caption{\label{fig:mu2X_A_Q2}
Ground-state (X) and excited-state (A) electronic matrix elements of 
$\hat{\mu}_x^2$,  $\hat{\mu}_y^2$ and $\hat{\mu}_z^2$ in atomic units, 
obtained at the RHF/6-31G* (X) and CIS/6-31G* (A) levels of theory along the
$\nu_2$ (C=O stretch) normal mode. All other $\hat{\mu}_k \hat{\mu}_l$ matrix elements
are zero (see Table \ref{tbl:grouptheory}).}
\end{figure}

\begin{figure}
\includegraphics[width=0.75\textwidth]{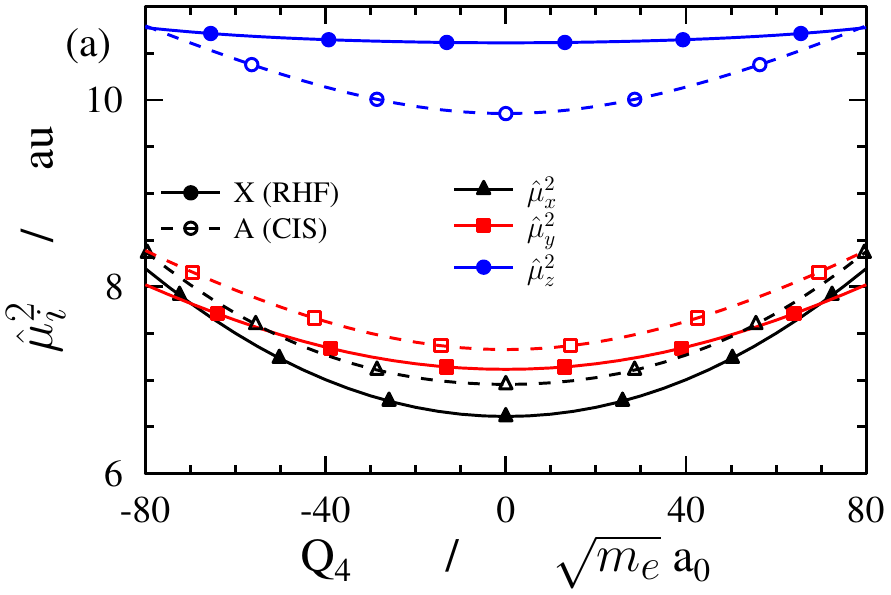}
\includegraphics[width=0.75\textwidth]{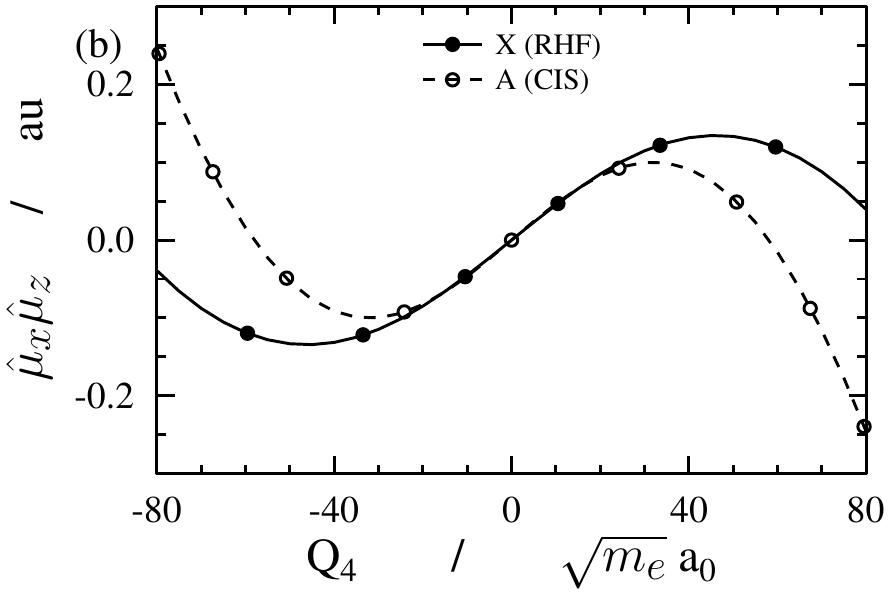}
\caption{\label{fig:mu2X_A_Q4}
Ground-state (X) and excited-state (A) electronic
matrix elements of $\hat{\mu}_x^2$, $\hat{\mu}_y^2$ and $\hat{\mu}_z^2$ (panel a),
and $\hat{\mu}_x\hat{\mu}_z$ (panel b) in atomic units, obtained at the RHF/6-31G* (X)
and CIS/6-31G* (A) levels of theory along the $\nu_4$ (out-of-plane) normal mode. 
All other $\hat{\mu}_k \hat{\mu}_l$ matrix elements are zero 
(see Table \ref{tbl:grouptheory}).}
\end{figure}

\begin{figure}
\includegraphics[width=0.6\textwidth]{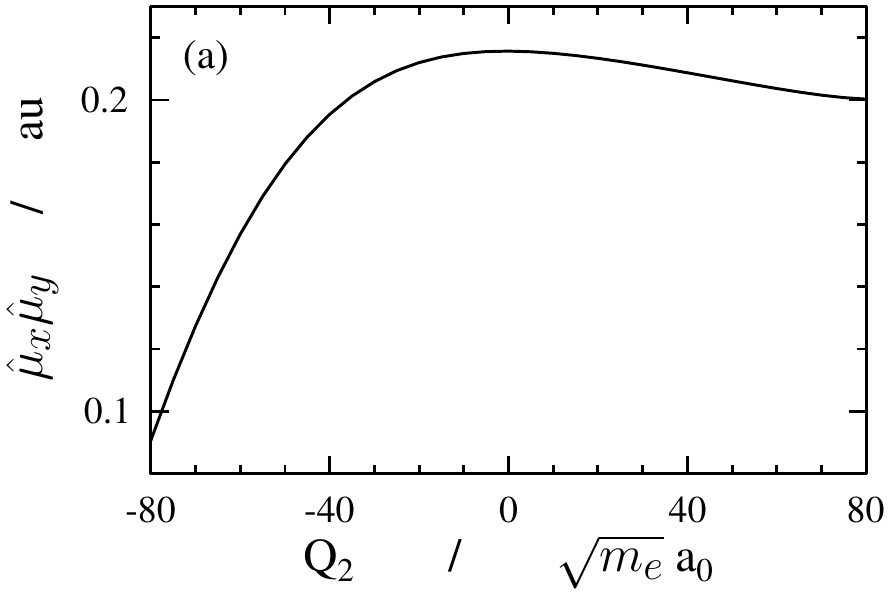}
\includegraphics[width=0.6\textwidth]{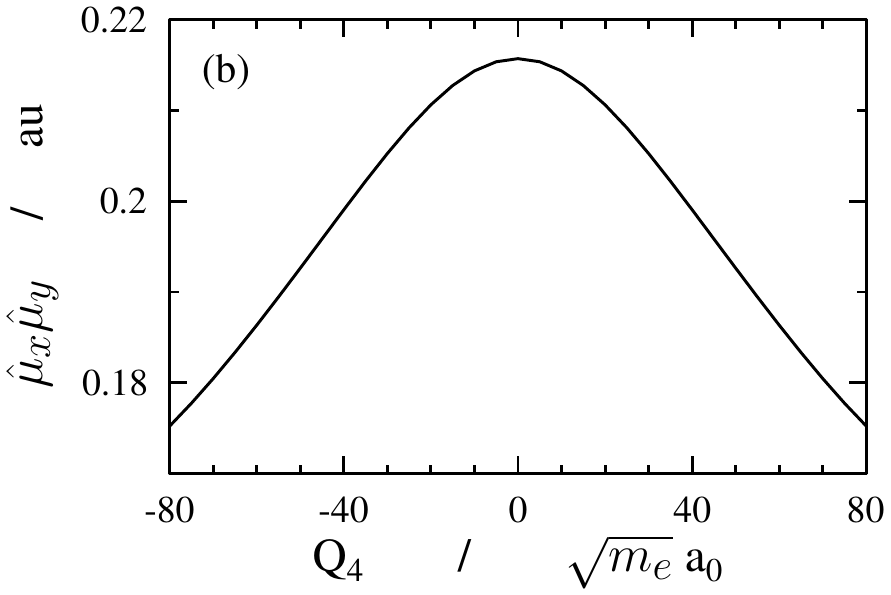}
\includegraphics[width=0.6\textwidth]{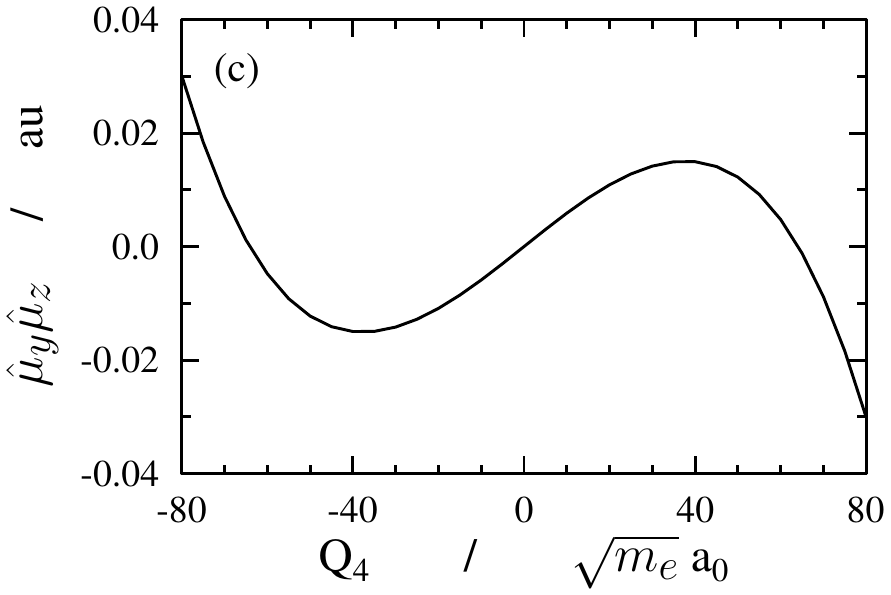}
\caption{\label{fig:mu2XA_Q2_Q4}
Electronic transition matrix elements of the operators $\hat{\mu}_x\hat{\mu}_y$ and
$\hat{\mu}_y\hat{\mu}_z$ between the ground (X) and excited (A) electronic states
in atomic units, obtained at the RHF/6-31G* (X) and CIS/6-31G* (A) levels of theory
(panels a and b:  $\langle\textrm{X}|\hat{\mu}_x\hat{\mu}_y|\textrm{A}\rangle$
along the $\nu_2$ (C=O stretch) and $\nu_4$ (out-of-plane) normal modes,
panel c: $\langle\textrm{X}|\hat{\mu}_y\hat{\mu}_z|\textrm{A}\rangle$ along the 
$\nu_4$ normal mode.) Note that all nonvanishing transition matrix elements are
shown (see Table \ref{tbl:grouptheory}).}
\end{figure}

This subsection discusses electronic structure results relevant for the DSE term.
Namely, ground-state and excited-state electronic matrix elements of the operators 
$\hat{\mu}_k \hat{\mu}_l$ as well as transition matrix elements
$\langle \textrm{X} | \hat{\mu}_k \hat{\mu}_l | \textrm{A} \rangle$ are presented.
Due to the high symmetry of the molecule, many matrix elements vanish and
in Figs. \ref{fig:mu2X_A_Q2}, \ref{fig:mu2X_A_Q4} and \ref{fig:mu2XA_Q2_Q4},
all nonzero $\hat{\mu}_k \hat{\mu}_l$ matrix elements are shown.
Note that numerical results (i.e., which $\hat{\mu}_k \hat{\mu}_l$ operators yield nonzero
matrix elements) are in line with group-theoretical considerations described
in Section \ref{sec:compdet} and Table \ref{tbl:grouptheory}.
Figs. \ref{fig:mu2X_A_Q2} and \ref{fig:mu2X_A_Q4} compare the ground-state (RHF/6-31G*)
and excited-state (CIS/6-31G*) matrix elements of the operators $\hat{\mu}_k \hat{\mu}_l$
for one-dimensional scans along the $\nu_2$ and $\nu_4$ vibrational modes, respectively.
In each case, rectilinear normal coordinates ($Q_2$ and $Q_4$) are used and the inactive 
normal coordinate is set to zero.
It is visible in Fig. \ref{fig:mu2X_A_Q4} that the only nonzero offdiagonal matrix element
($\hat{\mu}_x \hat{\mu}_z$) vanishes for $Q_4=0$ and its value is smaller than those of the
diagonal components $\hat{\mu}_k^2$ by an order of magnitude.
In addition, Fig. \ref{fig:mu2XA_Q2_Q4} presents nonzero transition matrix elements
of $\hat{\mu}_x \hat{\mu}_y$ and $\hat{\mu}_y \hat{\mu}_z$ between the ground and 
excited electronic states, again along the $\nu_2$ and $\nu_4$ modes.
In this case, all diagonal components $\hat{\mu}_k^2$ yield zero transition matrix elements.
Furthermore, the magnitudes of transition matrix elements shown in 
Fig. \ref{fig:mu2XA_Q2_Q4} are considerably smaller than those of the
ground-state and excited-state matrix elements of the diagonal components 
$\hat{\mu}_k^2$ (see Figs. \ref{fig:mu2X_A_Q2} and \ref{fig:mu2X_A_Q4}).

\subsection{Light-induced nonadiabatic effects}

\begin{figure}
\includegraphics[width=0.475\textwidth]{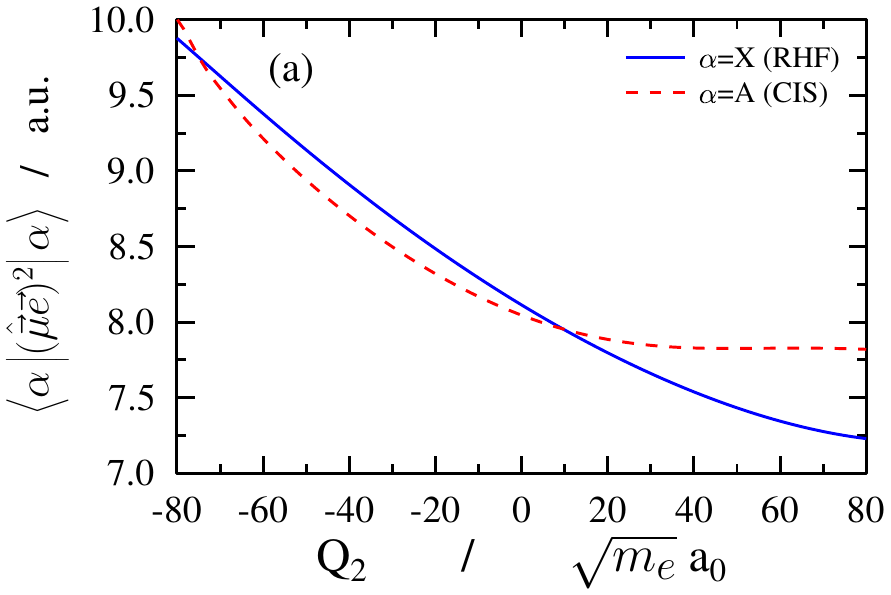}
\includegraphics[width=0.475\textwidth]{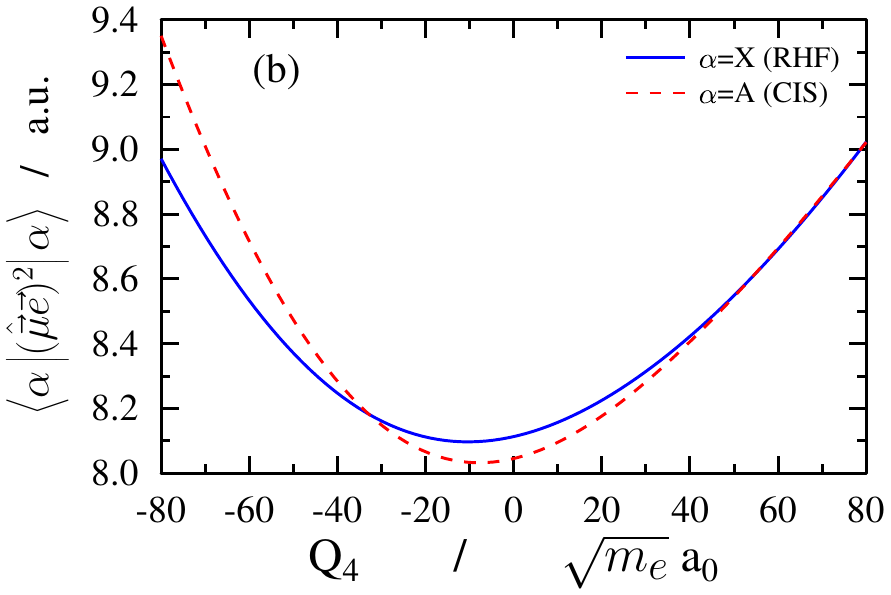}
\includegraphics[width=0.475\textwidth]{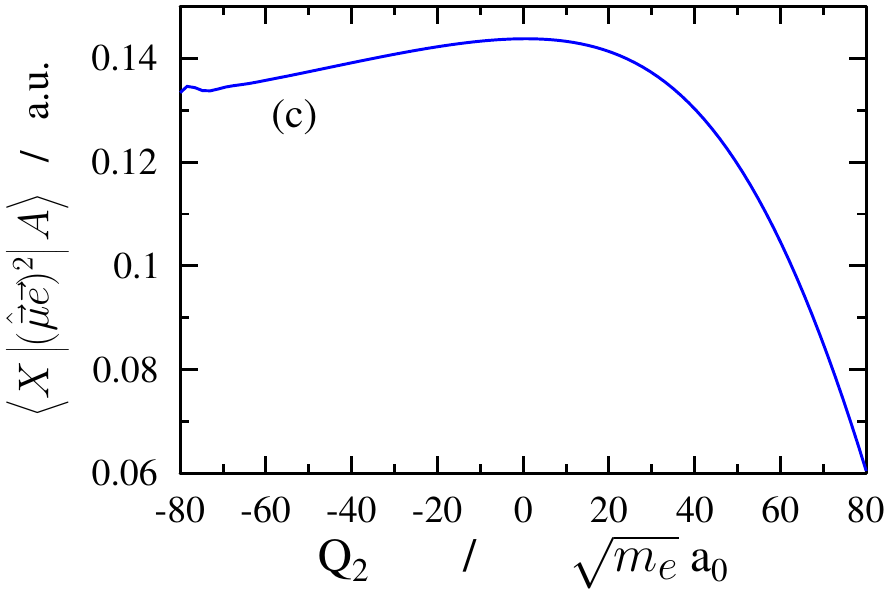}
\includegraphics[width=0.475\textwidth]{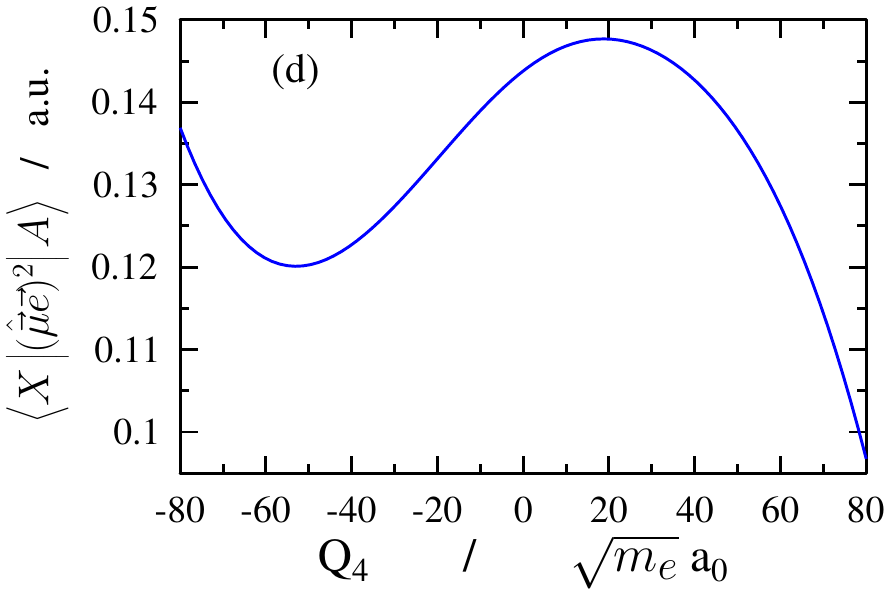}
\caption{\label{fig:dse_tot}
Ground-state (X) and excited-state (A) electronic matrix elements 
$\langle \alpha | (\hat{\vec{\mu}}\vec{e})^2 | \alpha \rangle$ 
($\alpha = \textrm{X}, \textrm{A}$) in atomic units, obtained at the RHF/6-31G* (X) 
and CIS/6-31G* (A) levels of theory along the  $\nu_2$ (C=O stretch, panel a) and 
$\nu_4$ (out-of-plane, panel b) normal modes. Electronic transition matrix elements
$\langle \textrm{X} | (\hat{\vec{\mu}}\vec{e})^2 | \textrm{A} \rangle$ are also shown
in atomic units along the $\nu_2$ (panel c) and $\nu_4$ (panel d) normal modes.
In all cases, the polarization vector is chosen as $\mathbf{e} = (1,1,1)/\sqrt{3}$.
$Q_2$ and $Q_4$ denote normal coordinates of the $\nu_2$ and $\nu_4$ normal modes,
respectively. Note that curves along $Q_4$ (panels b and d) are neither symmetric, 
nor antisymmetric under the transformation $Q_4 \rightarrow -Q_4$.}
\end{figure}

This subsection provides a discussion of how and to what extent LICIs are affected 
by the DSE. We set the cavity polarization to $\mathbf{e} = (1,1,1)/\sqrt{3}$ which 
assures that all nonzero TDM, PDM and DSE terms come into play.
Namely, in the 2D($\nu_2$,$\nu_4$) model, we get
\begin{equation}
    \langle \alpha | \hat{\vec{\mu}}\vec{e} | \alpha \rangle = 
    \frac{1}{\sqrt{3}} \sum_{k=x,z} \langle \alpha | \hat{\mu}_k | \alpha \rangle
\label{eq:pdm}
\end{equation}
and
\begin{equation}
    \langle \textrm{X} | \hat{\vec{\mu}}\vec{e} | \textrm{A} \rangle = 
    \frac{1}{\sqrt{3}} \langle \textrm{X} | \hat{\mu}_y | \textrm{A} \rangle
\label{eq:tdm}
\end{equation}
where $\alpha = \textrm{X}, \textrm{A}$ labels electronic states and the group-theoretical
results of Table \ref{tbl:grouptheory} can be used to assess that the $y$ component 
of the PDMs and the $x$ and $z$ components of the TDM vanish due to symmetry. 
Similarly, for the DSE, one can show that
\begin{equation}
    \langle \alpha | (\hat{\vec{\mu}}\vec{e})^2 | \alpha \rangle = 
    \frac{1}{3} \sum_{k=x,y,z} \langle \alpha | \hat{\mu}_k^2 | \alpha \rangle+
    \frac{2}{3} \langle \alpha | \hat{\mu}_x \hat{\mu}_z | \alpha \rangle
\label{eq:dseterm1}
\end{equation}
and
\begin{equation}
    \langle \textrm{X} | (\hat{\vec{\mu}}\vec{e})^2 | \textrm{A} \rangle = \frac{2}{3}
    (\langle \textrm{X} | \hat{\mu}_x \hat{\mu}_y | \textrm{A} \rangle + \langle \textrm{X} | \hat{\mu}_y \hat{\mu}_z | \textrm{A} \rangle).
\label{eq:dseterm2}
\end{equation}
One-dimensional scans for $\langle \alpha | (\hat{\vec{\mu}}\vec{e})^2 | \alpha \rangle$
and $\langle \textrm{X} | (\hat{\vec{\mu}}\vec{e})^2 | \textrm{A} \rangle$ along the
$\nu_2$ and $\nu_4$ normal modes are shown in Fig. \ref{fig:dse_tot}.

\begin{figure}
\includegraphics[width=0.85\textwidth]{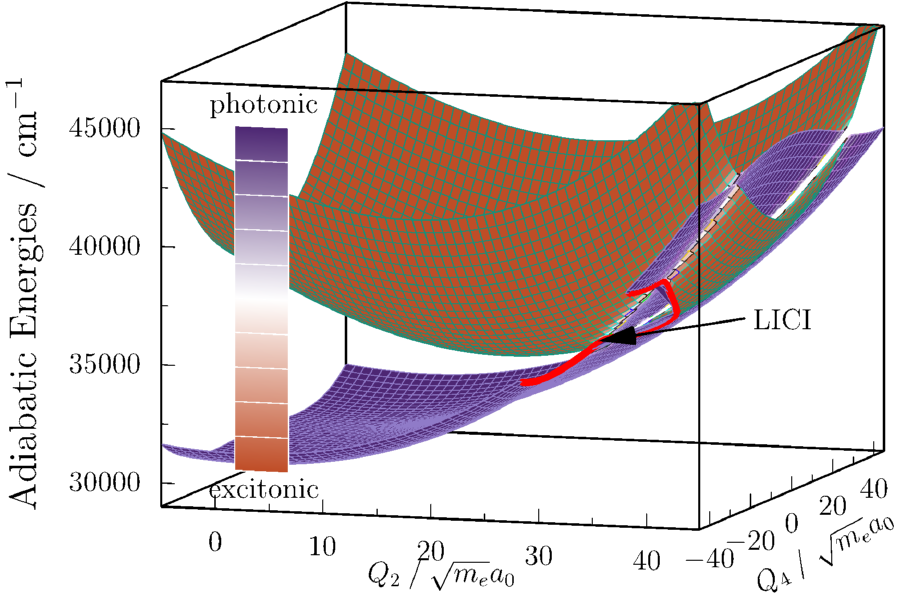}
\caption{\label{fig:popes}
Two-dimensional lower (1LP) and upper (1UP) polaritonic potential energy surfaces (PESs),
including the dipole self-energy term.
$Q_2$ and $Q_4$ denote normal coordinates of the $\nu_2$ (C=O stretch) and
$\nu_4$ (out-of-plane) normal modes.
The cavity wavenumber, coupling strength and polarization are chosen as
$\omega_\textrm{c}=34304.93~\textrm{cm}^{-1}$, $g=0.01~\textrm{au}$ and
$\mathbf{e} = (1,1,1)/\sqrt{3}$, respectively.
The character of the polaritonic PESs is indicated by different colors 
(purple: photonic, orange: excitonic).
The light-induced conical intersection (LICI, located at 
$Q_2=27.3724 \sqrt{m_\textrm{e}} a_0$ and $Q_4=0.0179 \sqrt{m_\textrm{e}} a_0$) 
between the 1LP and 1UP PESs is explicitly marked in the figure.
Polaritonic PESs are referenced to the minimum of the ground (lowest) polaritonic PES. 
A half circle (shown in red) is cut from the 1LP PES in order to make the LICI visible.}
\end{figure}

\begin{figure}
\includegraphics[width=0.475\textwidth]{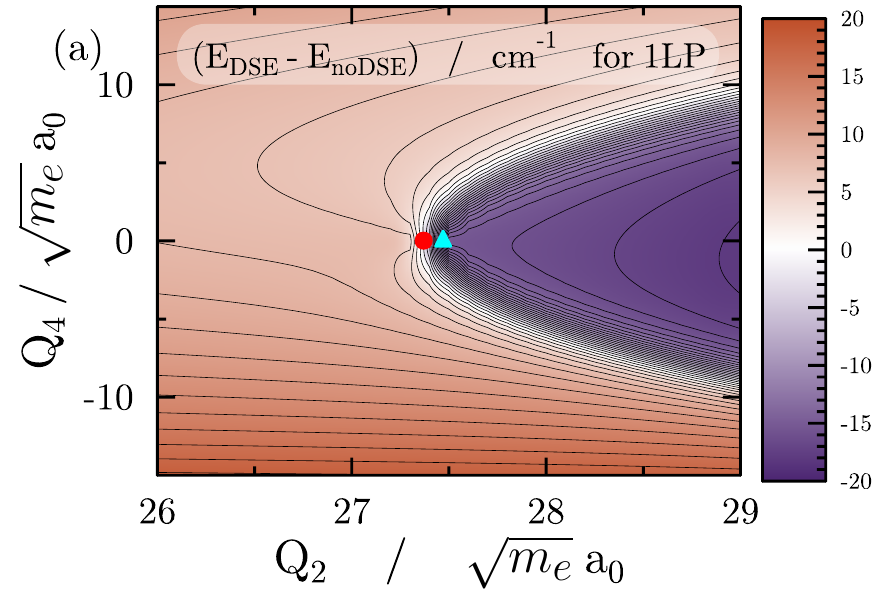}
\includegraphics[width=0.475\textwidth]{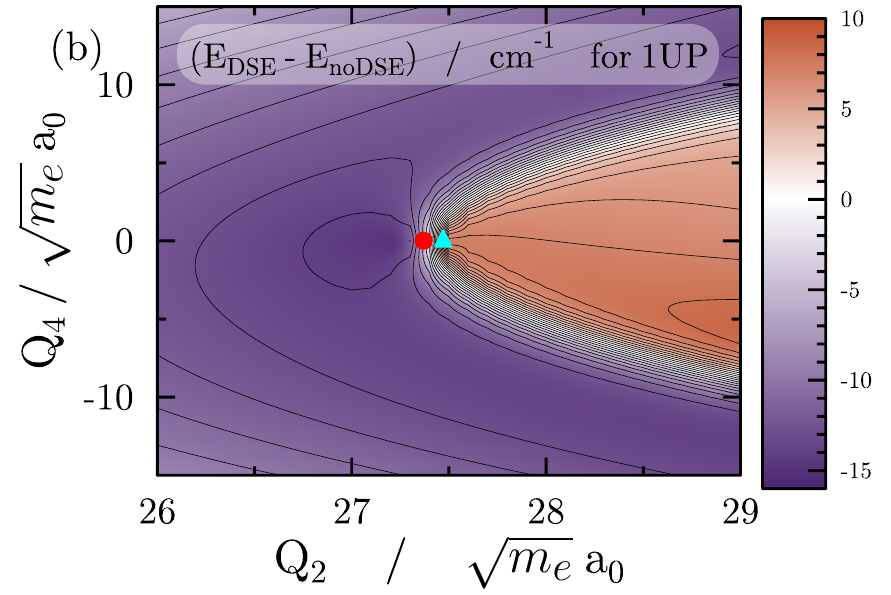}
\caption{\label{fig:diffpopes}
Differences between polaritonic potential energy surfaces (PESs) obtained with and without
the dipole self-energy (DSE) term for the 1LP (panel a) and 1UP (panel b) polaritonic 
states. $Q_2$ and $Q_4$ denote normal coordinates of the $\nu_2$ (C=O stretch)
and $\nu_4$ (out-of-plane) normal modes.
The cavity wavenumber, coupling strength and polarization are chosen as
$\omega_\textrm{c}=34304.93~\textrm{cm}^{-1}$, $g=0.01~\textrm{au}$ and
$\mathbf{e} = (1,1,1)/\sqrt{3}$, respectively.
Positions of the light-induced conical intersection between the 1LP and 1UP polaritonic 
PESs are explicitly marked by the symbols $\blacktriangle$ (without DSE, 
$Q_2=27.4745 \sqrt{m_\textrm{e}} a_0$ and $Q_4=0$) and $\bullet$ 
(with DSE, $Q_2=27.3724 \sqrt{m_\textrm{e}} a_0$ and $Q_4=0.0179 \sqrt{m_\textrm{e}} a_0$).
Polaritonic PESs are referenced to the minimum of the ground (lowest) polaritonic PES.}
\end{figure}

Polaritonic PESs can be constructed by diagonalizing the potential energy part of
the cavity-molecule Hamiltonian of Eq. \eqref{eq:cavity_H}, either with or without the 
DSE term. Fig. \ref{fig:popes} shows the 1LP and 1UP polaritonic PESs for 
$\omega_\textrm{c}=34304.93~\textrm{cm}^{-1}$, $g=0.01~\textrm{au}$ and
$\mathbf{e} = (1,1,1)/\sqrt{3}$ with the DSE included in the potential energy matrix.
We note that $g$ has the dimension of  electric field and the coupling energy
can be given as the product of $g$ and the molecular dipole moment. Since the TDM varies 
between $-0.2~\textrm{au}$ and $0.2~\textrm{au}$ in the nuclear configuration space relevant
for the current results, the coupling energy between electronic states X and A does not 
exceed  $\sim 500~\textrm{cm}^{-1}$.
It is visible in Fig. \ref{fig:popes} that the 1LP and 1UP PES form a LICI, which is also 
the case if the DSE is omitted from the potential energy matrix. 
However, the DSE modifies the position of the LICI. Namely, without DSE, the LICI is
situated at $Q_2=27.4745 \sqrt{m_\textrm{e}} a_0$ and $Q_4=0$, 
which is shifted to $Q_2=27.3724 \sqrt{m_\textrm{e}} a_0$ and
$Q_4=0.0179 \sqrt{m_\textrm{e}} a_0$ if the DSE is taken into account. 
The energetic position of the LICI  corresponds to $34826.86~\textrm{cm}^{-1}$ 
without DSE and $34797.91~\textrm{cm}^{-1}$ with DSE,
both referenced to the minimum of the respective ground (lowest) polaritonic PES.
LICI positions are explicitly marked in Fig. \ref{fig:diffpopes} where differences 
between polaritonic PESs obtained with and without the DSE term are shown for the 
1LP and 1UP polaritonic states. 

Although 
$V_\textrm{X}(Q_2,-Q_4)=V_\textrm{X}(Q_2,Q_4)$ and
$V_\textrm{A}(Q_2,-Q_4)=V_\textrm{A}(Q_2,Q_4)$,
the ``difference polaritonic'' PESs in  Fig. \ref{fig:diffpopes} are not symmetric under the transformation 
$Q_4 \rightarrow -Q_4$. 
The same observation applies to the polaritonic PESs themselves, which can be understood by 
inspecting the symmetry of the TDM, PDM and DSE terms. Using the data given in
Table \ref{tbl:grouptheory}, one can determine that the $z$ component of the PDM
is symmetric while the $x$ component of the PDM and the $y$ component of the TDM are 
antisymmetric under $Q_4 \rightarrow -Q_4$.
Therefore, the expression for $\langle \alpha | \hat{\vec{\mu}}\vec{e} | \alpha \rangle$
in Eq. \eqref{eq:pdm} is neither symmetric nor antisymmetric, 
which breaks the symmetry of the resulting polaritonic PESs.
However, if the DSE is omitted, the LICI is still located at a symmetric position 
with $Q_4=0$. The DSE terms in Eqs. \eqref{eq:dseterm1} and \eqref{eq:dseterm2} 
induce further symmetry breaking, which is also
visible in panels b and d of Fig. \ref{fig:dse_tot} (that is, one-dimensional 
$\langle \alpha | (\hat{\vec{\mu}}\vec{e})^2 | \alpha \rangle$ and
$\langle \textrm{X} | (\hat{\vec{\mu}}\vec{e})^2 | \textrm{A} \rangle$ curves as a 
function of $Q_4$ are neither symmetric, nor antisymmetric under $Q_4 \rightarrow -Q_4$).
As a consequence, the position of the LICI is shifted to $Q_4 \ne 0$ if the DSE is 
included in the potential energy matrix.

\subsection{Time-dependent quantum dynamics}

In this subsection, differences between time-dependent quantum-dynamical results obtained
with and without the DSE term are discussed. Two different cavity wavenumber values of 
$\omega_\textrm{c}=34304.93~\textrm{cm}^{-1}$ and
$\omega_\textrm{c}=34370.27~\textrm{cm}^{-1}$ 
are employed, while the other cavity parameters are set to $g=0.01~\textrm{au}$, 
$\gamma_\textrm{c}=10^{-4}~\textrm{au}$ (equivalent to a lifetime of 
$\tau=\hbar/\gamma_\textrm{c} = 241.9 ~\textrm{fs}$) and $\mathbf{e} = (1,1,1)/\sqrt{3}$.
The molecule coupled to a single cavity mode is pumped with a laser pulse,
the laser intensity equals $I = 5~\textrm{TW/cm}^2$ and the pulse length is chosen as 
$T = 200~\textrm{fs}$ (see the text under Eq. \eqref{eq:Schrodinger} for the definition 
of the envelope function). The field amplitude $E_0$ is related to the intensity by 
the formula $E_0 = \sqrt{2I/(c\epsilon_0)}$
($c$ is the speed of light in vacuum and $\epsilon_0$ is the vacuum permittivity).

\begin{figure}
\includegraphics[width=0.475\textwidth]{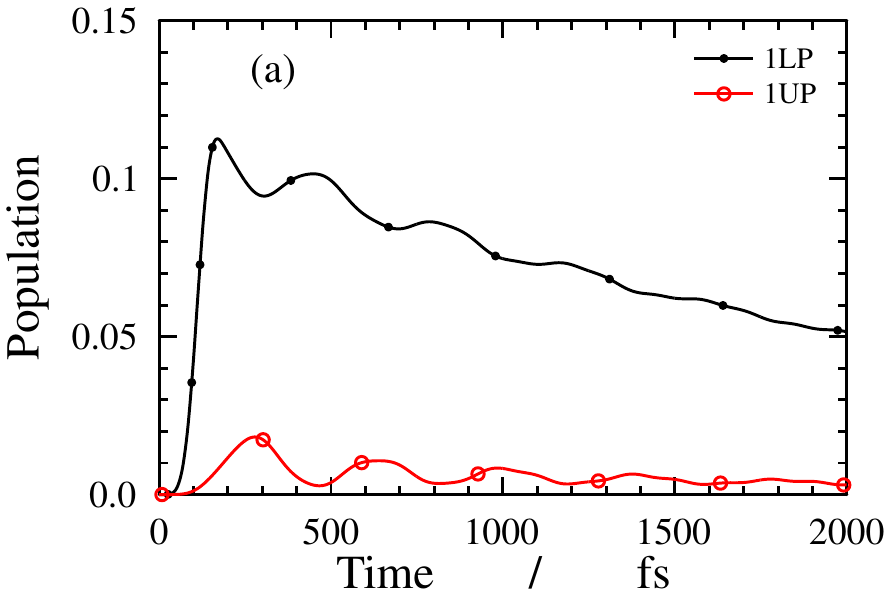}
\includegraphics[width=0.475\textwidth]{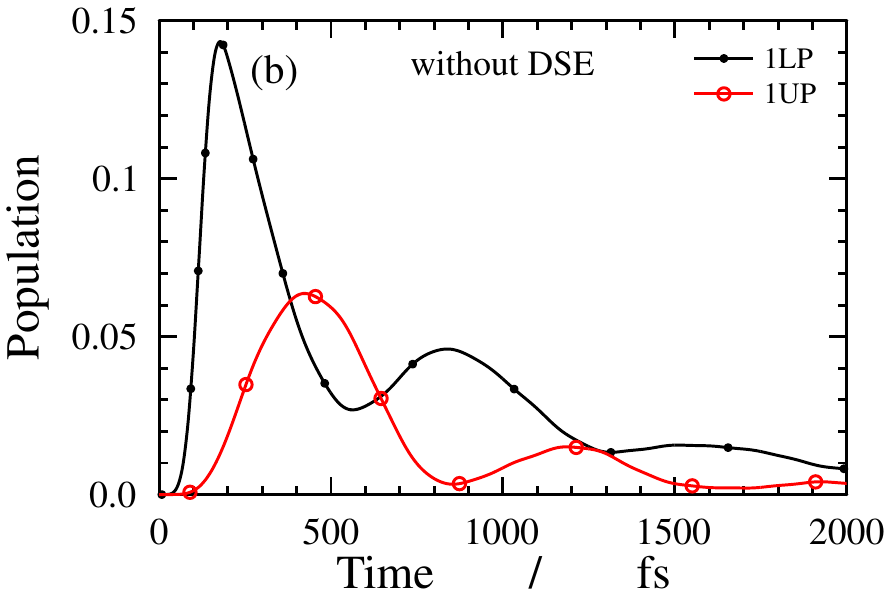}
\includegraphics[width=0.475\textwidth]{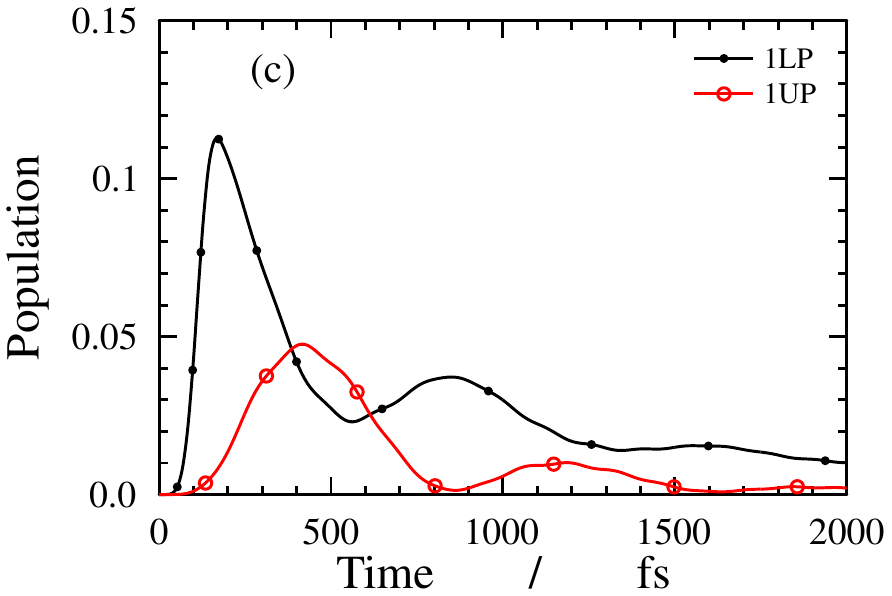}
\includegraphics[width=0.475\textwidth]{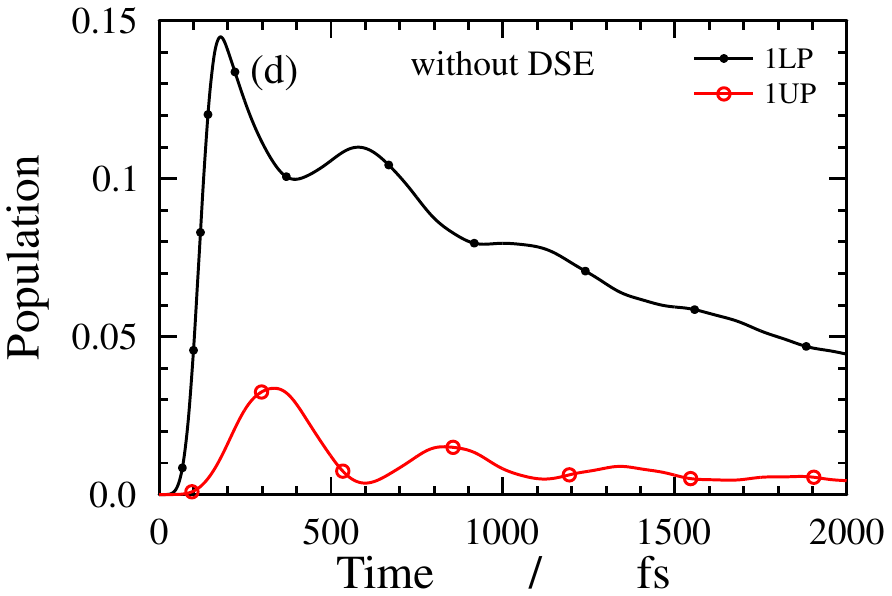}
\caption{\label{fig:ad_pop_1}
Populations of polaritonic states (1LP and 1UP) as a function of time obtained with (panels a and c) and without (panels b and d) the DSE term. The cavity wavenumber is chosen as 
$\omega_\textrm{c}=34304.93~\textrm{cm}^{-1}$ (panels a and b) and 
$\omega_\textrm{c}=34370.27~\textrm{cm}^{-1}$ (panels c and d).
Other cavity and laser parameters equal $g=0.01~\textrm{au}$, $\gamma_\textrm{c}=10^{-4}~\textrm{au}$, $\mathbf{e} = (1,1,1)/\sqrt{3}$, 
$\omega=34300~\textrm{cm}^{-1}$, $I = 5~\textrm{TW/cm}^2$ and $T = 200~\textrm{fs}$.
Note the sensitivity of 1LP and 1UP populations to $\omega_\textrm{c}$ and
the similarity between panels a and d, and panels b and c.}
\end{figure}

\begin{figure}
\includegraphics[width=0.475\textwidth]{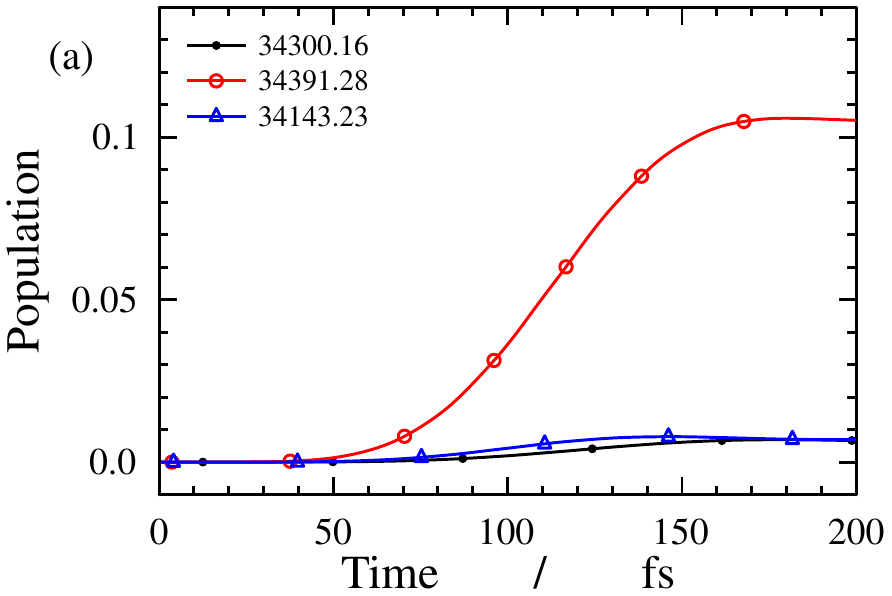}
\includegraphics[width=0.475\textwidth]{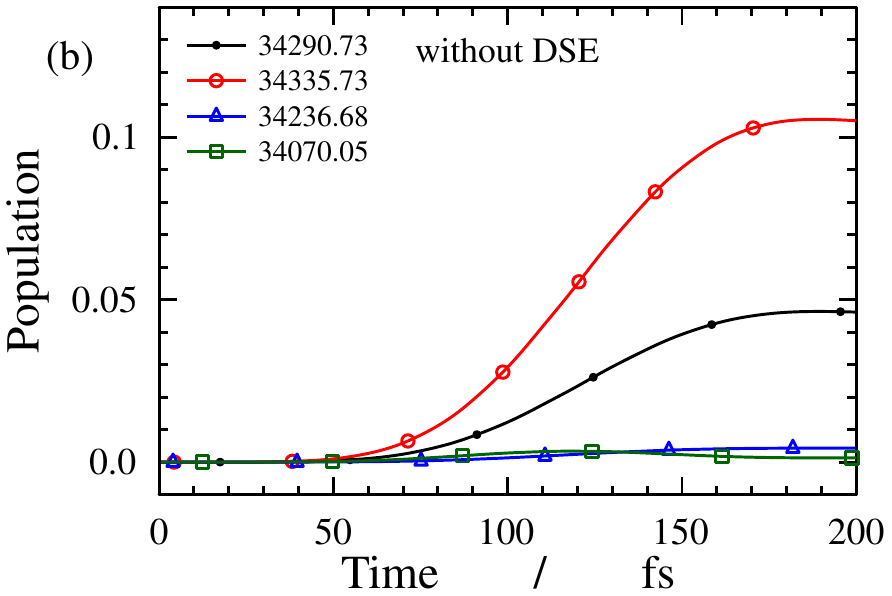}
\includegraphics[width=0.475\textwidth]{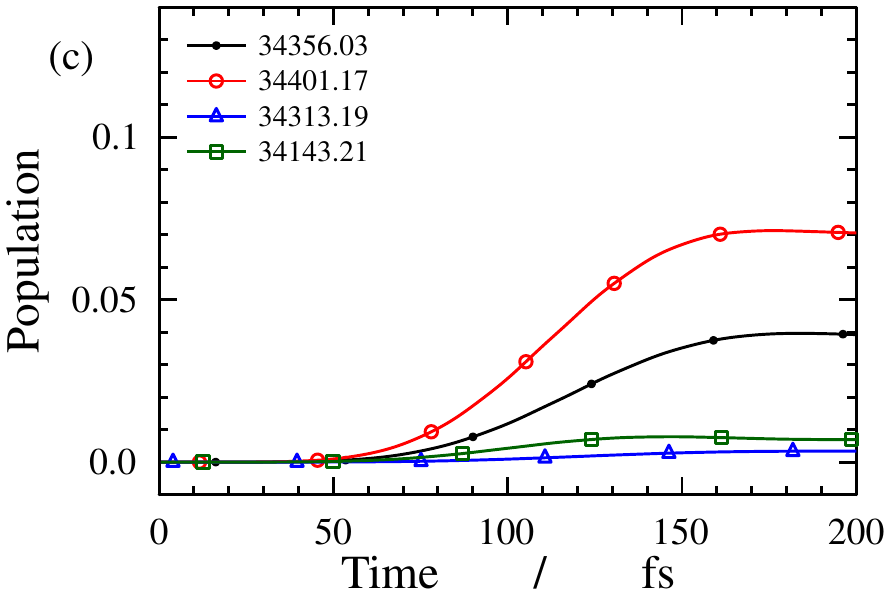}
\includegraphics[width=0.475\textwidth]{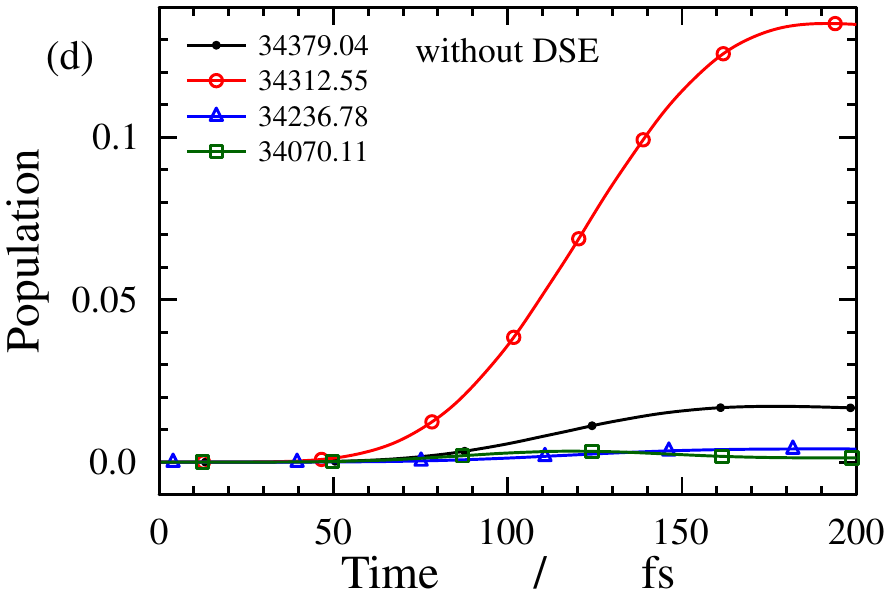}
\caption{\label{fig:ed_1}
Populations of cavity-molecule eigenstates populated by laser excitation
as a function of time (panels a and c: with DSE, panels b and d: without DSE).
The cavity wavenumber is chosen as 
$\omega_\textrm{c}=34304.93~\textrm{cm}^{-1}$ (panels a and b) and 
$\omega_\textrm{c}=34370.27~\textrm{cm}^{-1}$ (panels c and d).
Other cavity and laser parameters equal $g=0.01~\textrm{au}$, $\gamma_\textrm{c}=10^{-4}~\textrm{au}$, $\mathbf{e} = (1,1,1)/\sqrt{3}$, 
$\omega=34300~\textrm{cm}^{-1}$, $I = 5~\textrm{TW/cm}^2$ and $T = 200~\textrm{fs}$.
Energy levels in the plot legends are given in units of $\textrm{cm}^{-1}$ and referenced
to the lowest energy level.
Note the sensitivity of eigenstate populations to $\omega_\textrm{c}$ and
the similarity between panels a and d, and panels b and c.}
\end{figure}

First, the laser wavenumber is set to $\omega=34300~\textrm{cm}^{-1}$. 
Fig. \ref{fig:ad_pop_1} shows time-dependent populations of the 1LP and 1UP polaritonic
states during and after laser excitation. Panels a and c, and panels b and d present 
results with and without the DSE term, respectively. Data depicted in panels a and b of
Fig. \ref{fig:ad_pop_1} were obtained with $\omega_\textrm{c}=34304.93~\textrm{cm}^{-1}$,
while panels c and d correspond to $\omega_\textrm{c}=34370.27~\textrm{cm}^{-1}$
(these choices for $\omega_\textrm{c}$ will be justified later).
It is easy to notice the similarity between panels a and d, and panels b and c of
Fig. \ref{fig:ad_pop_1}. In addition, even though the DSE adds relatively small shifts
to the energy levels and potentials, panels of Fig. \ref{fig:ad_pop_1} also show
how sensitive the dynamics is to the value of $\omega_\textrm{c}$ and whether the DSE
is included in the cavity-molecule Hamiltonian.
Similar conclusions can be drawn by inspecting Fig. \ref{fig:ed_1} (following the
structure of Fig. \ref{fig:ad_pop_1}) where populations of cavity-molecule eigenstates
relevant for the dynamics are shown between $0~\textrm{fs}$ and $T = 200~\textrm{fs}$.
If the DSE term is taken into account and $\omega_\textrm{c}=34304.93~\textrm{cm}^{-1}$,
dominantly one eigenstate ($34391.28~\textrm{cm}^{-1}$, referenced to the lowest energy
level of $1526.94~\textrm{cm}^{-1}$, see panel a) is populated by the laser pulse, 
while without the DSE the population is mainly transferred to two close-lying eigenstates 
($34290.73~\textrm{cm}^{-1}$ and $34335.73~\textrm{cm}^{-1}$, both referenced to the 
lowest energy level of $1522.71~\textrm{cm}^{-1}$, see panel b) with 
$\omega_\textrm{c}=34304.93~\textrm{cm}^{-1}$.
The latter two eigenstates can be characterized as superpositions of the X vibrational
ground state ($1521.48~\textrm{cm}^{-1}$, dressed with one photon) and an 
excited A vibrational eigenstate ($35826.42~\textrm{cm}^{-1}$, dressed with zero photons)
which are resonantly coupled by the cavity mode.
If the DSE term is included in the Hamiltonian, the cavity frequency is no longer 
resonant with any of the $\textrm{X} \rightarrow \textrm{A}$ transitions and the 
dominantly-populated cavity-molecule eigenstate can be well approximated as an
A vibrational eigenstate with zero photons.

If the DSE is taken into account, the Hamiltonians
\begin{equation}
    \hat{H}_\alpha = \hat{T}+\hat{V}_\alpha+\frac{g^2}{\hbar\omega_\textrm{c}} 
        \hat{D}_\alpha
\end{equation}
appearing on the diagonal of the matrix Hamiltonian of Eq. \eqref{eq:cavity_H}
are central to determining conditions for resonant cavity-molecule coupling. Here,
$\hat{D}_\alpha = \langle \alpha |(\vec{\mu}\vec{e})^2| \alpha \rangle$ with 
$\alpha = \textrm{X}, \textrm{A}$, and
the ground-state and excited-state PESs are shifted by the DSE terms proportional 
to $\hat{D}_\alpha$. As a consequence, the effective  X and A vibrational
energy levels become dependent on cavity parameters, which renders the condition for 
resonant cavity-molecule coupling less obvious than in the no DSE case where the
resonance condition is that levels of X shifted by $\hbar \omega_\textrm{c}$ coincide 
with levels of A. One can apply first-order perturbation theory to approximate the 
energy levels of $\hat{H}_\alpha$, which yields
\begin{equation}
    E_{\alpha,i} = E_{\alpha,i}^{(0)}+\frac{g^2}{\hbar\omega_\textrm{c}}
        \langle \psi_{\alpha,i}^{(0)} | \hat{D}_\alpha | \psi_{\alpha,i}^{(0)} \rangle
\end{equation}
where zeroth-order energy levels and eigenstates are defined in the absence of DSE by
\begin{equation}
    \hat{H}_\alpha^{(0)} | \psi_{\alpha,i}^{(0)} \rangle = 
        E_{\alpha,i}^{(0)} | \psi_{\alpha,i}^{(0)} \rangle
\end{equation}
with $\hat{H}_\alpha^{(0)} = \hat{T}+\hat{V}_\alpha$.
With this, the resonance condition can be readily formulated as
\begin{equation}
    E_{\textrm{A},j}-E_{\textrm{X},i} = E_{\textrm{A},j}^{(0)}-E_{\textrm{X},i}^{(0)}+
    \frac{g^2}{\hbar\omega_\textrm{c}}
    (\langle \psi_{\textrm{A},j}^{(0)} | \hat{D}_\textrm{A} | \psi_{\textrm{A},j}^{(0)} \rangle-
    \langle \psi_{\textrm{X},i}^{(0)} | \hat{D}_\textrm{X} | \psi_{\textrm{X},i}^{(0)} \rangle)=
    \hbar \omega_\textrm{c}
  \label{eq:resonance}
\end{equation}
which can be solved for the resonance frequency $\omega_\textrm{c}$.

Using Eq. \eqref{eq:resonance} and choosing 
$E_{\textrm{A},j}^{(0)}=35826.42~\textrm{cm}^{-1}$ and
$E_{\textrm{X},i}^{(0)} = 1521.48~\textrm{cm}^{-1}$ 
(these are the vibrational levels that are resonantly coupled by the cavity for
$\omega_\textrm{c}=34304.93~\textrm{cm}^{-1}$ if the DSE is omitted), 
the resonant cavity wavenumber becomes
$\omega_\textrm{c}=34370.27~\textrm{cm}^{-1}$ for the DSE case. 
Panels c and d of Fig. \ref{fig:ad_pop_1} show 1LP and 1UP populations (panel c:
with DSE, panel d: without DSE) for $\omega_\textrm{c}=34370.27~\textrm{cm}^{-1}$.
Now, due to the  modified cavity wavenumber, cavity-molecule resonance occurs if the 
DSE term is included and resonance is disrupted if the DSE is neglected.
Populations of relevant cavity-molecule eigenstates shown in panels c and d of 
Fig. \ref{fig:ed_1} further support the conclusions drawn here.
The analysis of resonance conditions also clarifies why panels a and d, and panels
b and c are similar to each other in Figs. \ref{fig:ad_pop_1} and \ref{fig:ed_1}.
Namely, in both figures, panels a and d correspond to the nonresonant DSE and no DSE
cases, while panels b and c pertain to resonant cavity-molecule coupling.

\begin{figure}
\includegraphics[width=0.475\textwidth]{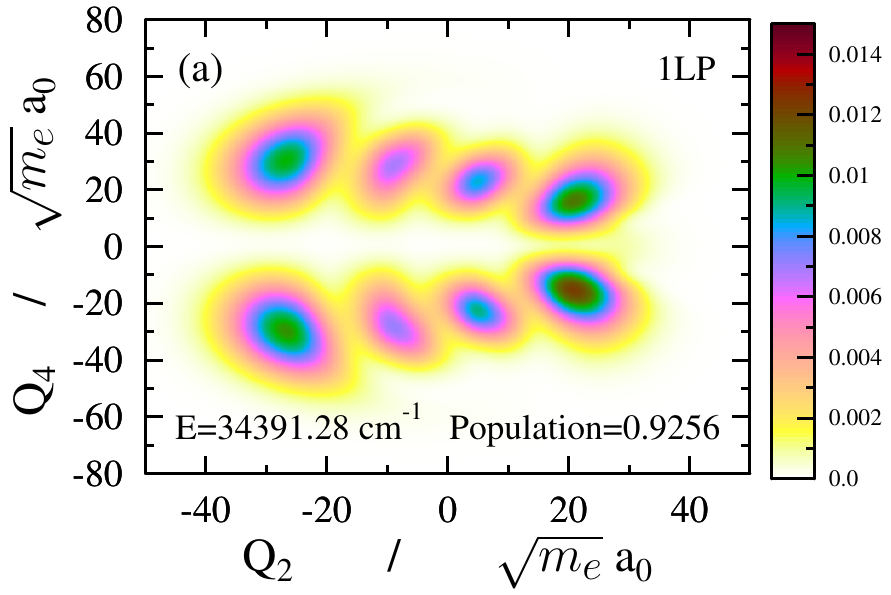}
\includegraphics[width=0.475\textwidth]{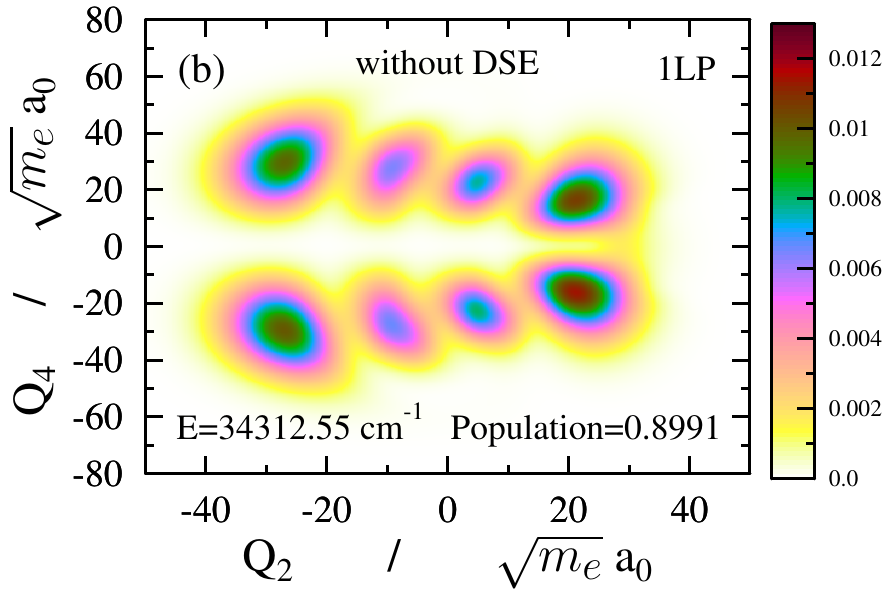}
\includegraphics[width=0.475\textwidth]{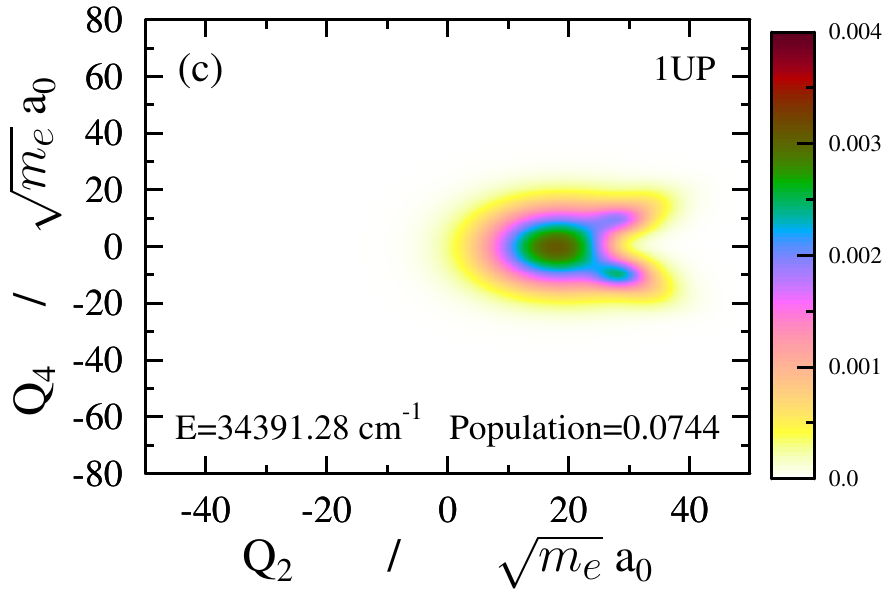}
\includegraphics[width=0.475\textwidth]{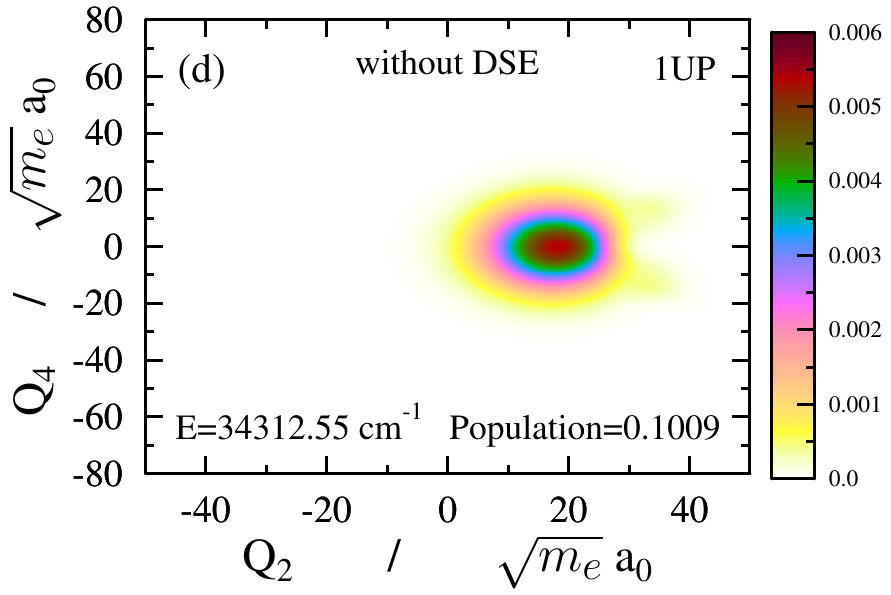}
\caption{\label{fig:dens_11}
Probability densities (1LP and 1UP states) for  cavity-molecule eigenstates 
populated by laser excitation (panels a and c: with DSE, panels b and d: without DSE).
The cavity wavenumber is chosen as $\omega_\textrm{c}=34304.93~\textrm{cm}^{-1}$ 
(panels a and c) and $\omega_\textrm{c}=34370.27~\textrm{cm}^{-1}$ (panels b and d).
Other cavity parameters equal $g=0.01~\textrm{au}$ and $\mathbf{e} = (1,1,1)/\sqrt{3}$.
$Q_2$ and $Q_4$ denote normal coordinates of the $\nu_2$ (C=O stretch) 
and $\nu_4$ (out-of-plane) normal modes.
Energy levels (given in the figures) are referenced to the lowest energy level.
Populations of the 1LP and 1UP states are also given in each panel.
Note the similarity between panel pairs in each row.}
\end{figure}

\begin{figure}
\includegraphics[width=0.425\textwidth]{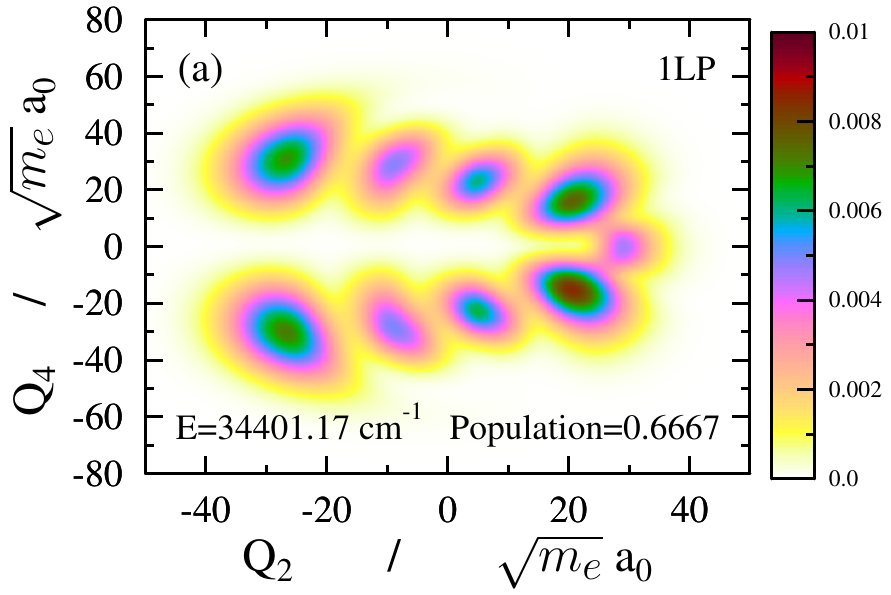}
\includegraphics[width=0.425\textwidth]{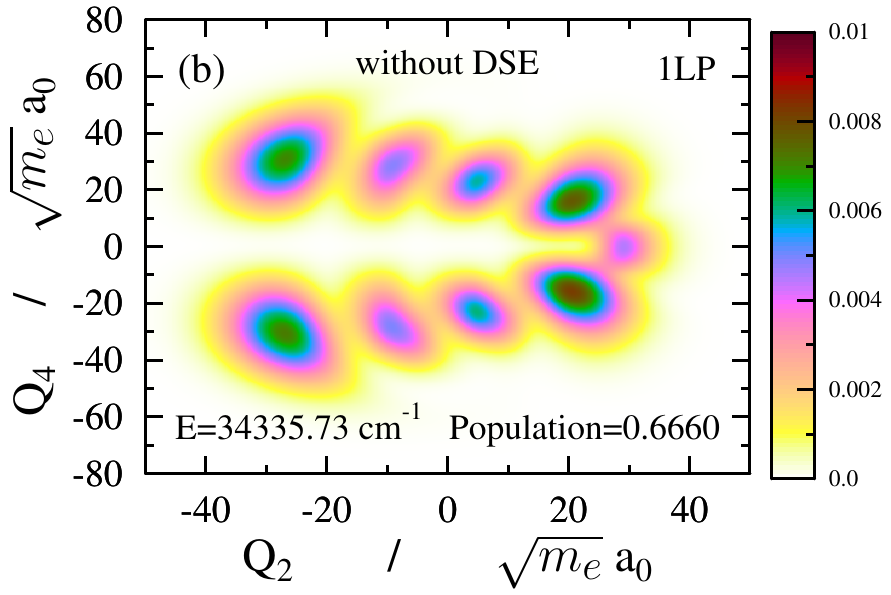}
\includegraphics[width=0.425\textwidth]{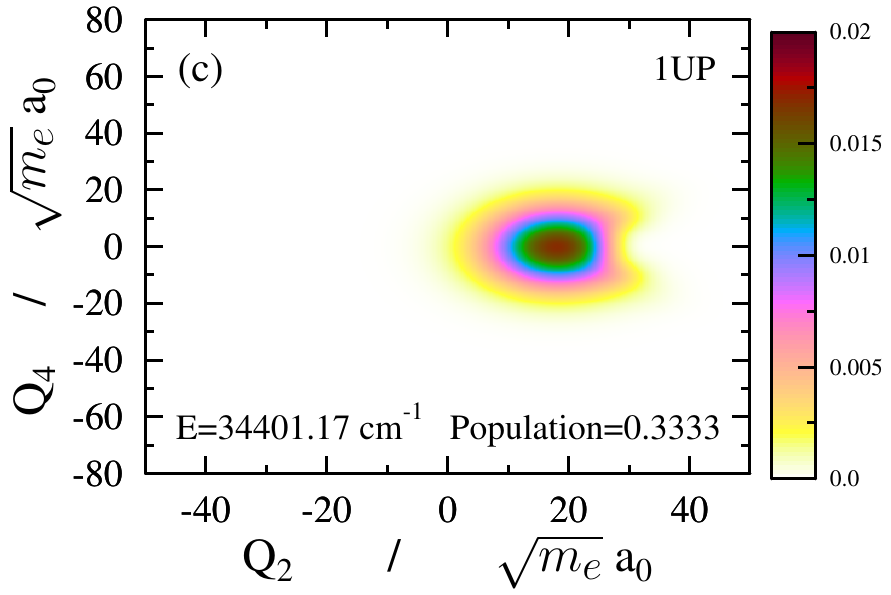}
\includegraphics[width=0.425\textwidth]{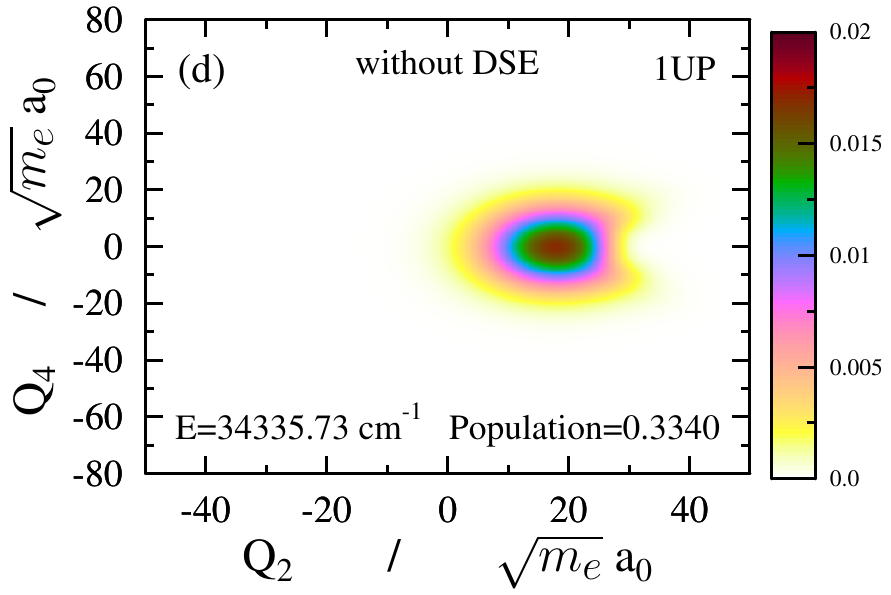}
\includegraphics[width=0.425\textwidth]{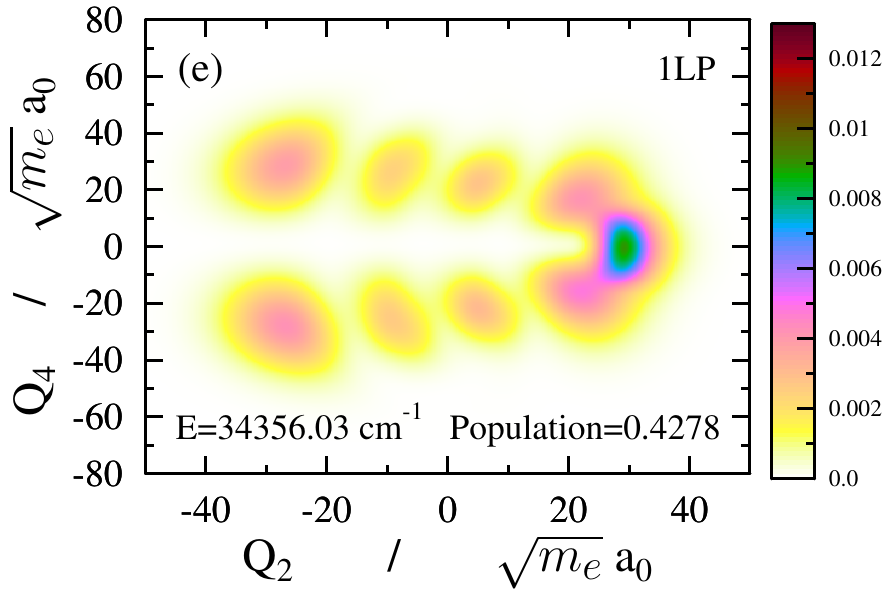}
\includegraphics[width=0.425\textwidth]{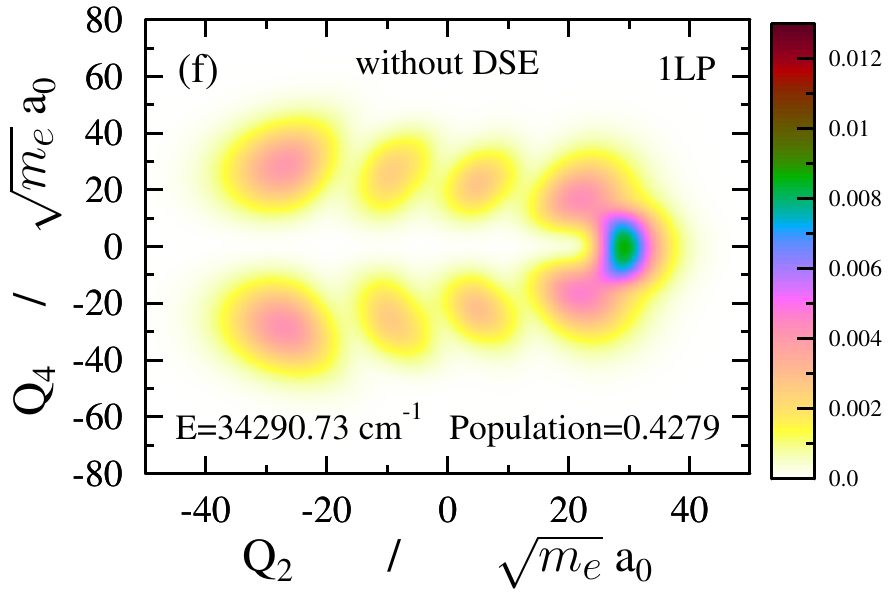}
\includegraphics[width=0.425\textwidth]{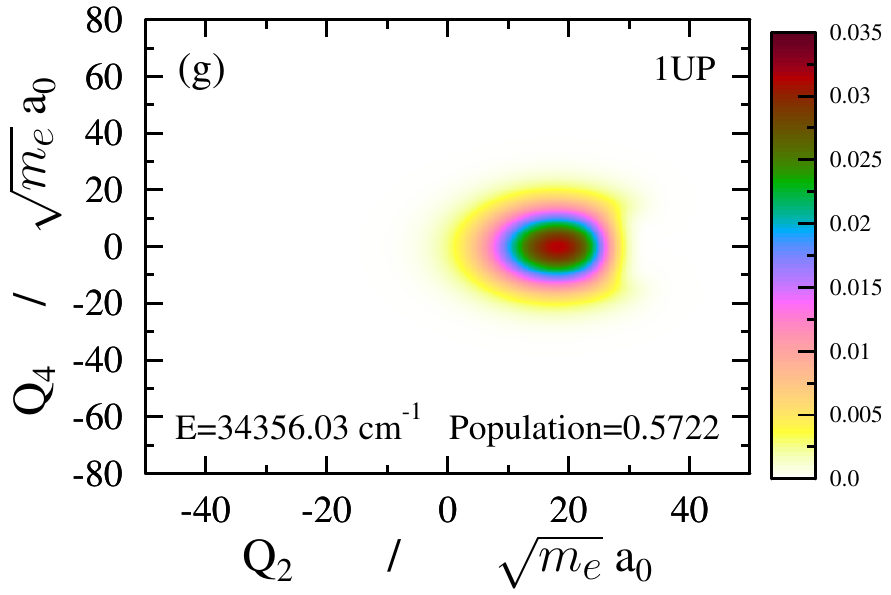}
\includegraphics[width=0.425\textwidth]{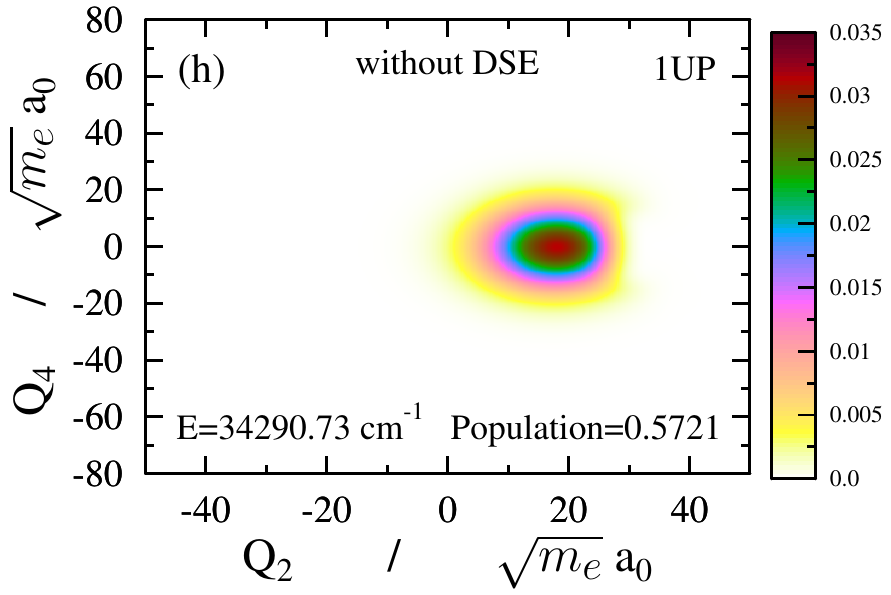}
\caption{\label{fig:dens_12}
Probability densities (1LP and 1UP states) for  cavity-molecule eigenstates 
populated by laser excitation (panels a, c, e and g: with DSE, panels b, d, f and h: without DSE).
The cavity wavenumber is chosen as $\omega_\textrm{c}=34370.27~\textrm{cm}^{-1}$
(panels a, c, e and g) and $\omega_\textrm{c}=34304.93~\textrm{cm}^{-1}$ 
(panels b, d, f and h).
Other cavity parameters equal $g=0.01~\textrm{au}$ and $\mathbf{e} = (1,1,1)/\sqrt{3}$.
$Q_2$ and $Q_4$ denote normal coordinates of the $\nu_2$ (C=O stretch) 
and $\nu_4$ (out-of-plane) normal modes.
Energy levels (given in the figures) are referenced to the lowest energy level.
Populations of the 1LP and 1UP states are also given in each panel.
Note the similarity between panel pairs in each row.}
\end{figure}

Figs. \ref{fig:dens_11} and \ref{fig:dens_12} depict 1LP and 1UP probability 
densities for  cavity-molecule eigenstates relevant for the interpretation of the 
current results.
It can be seen that one can establish a one-to-one correspondence between eigenstates
obtained with and without the DSE term. 
Namely, Fig. \ref{fig:dens_11} compares 1LP and 1UP probability densities for the
eigenstates $34391.28~\textrm{cm}^{-1}$ (with DSE) and $34312.55~\textrm{cm}^{-1}$
(without DSE), both of which correspond to the dominantly-populated eigenstates in
the nonresonant cases of Fig. \ref{fig:ed_1}.
Moreover, Fig. \ref{fig:dens_12} relates relevant eigenstates of the resonant DSE and
no DSE cases ($34401.17~\textrm{cm}^{-1}$ with DSE and $34335.73~\textrm{cm}^{-1}$ 
without DSE, $34356.03~\textrm{cm}^{-1}$ with DSE and $34290.73~\textrm{cm}^{-1}$ 
without DSE). For all eigenstate pairs, the respective probability densities agree well.
It is also clear from the results that the DSE perturbs the relevant energy levels by 
less than $100~\textrm{cm}^{-1}$ (when referenced to the lowest energy level) in this
particular case.

\begin{figure}
\includegraphics[width=0.475\textwidth]{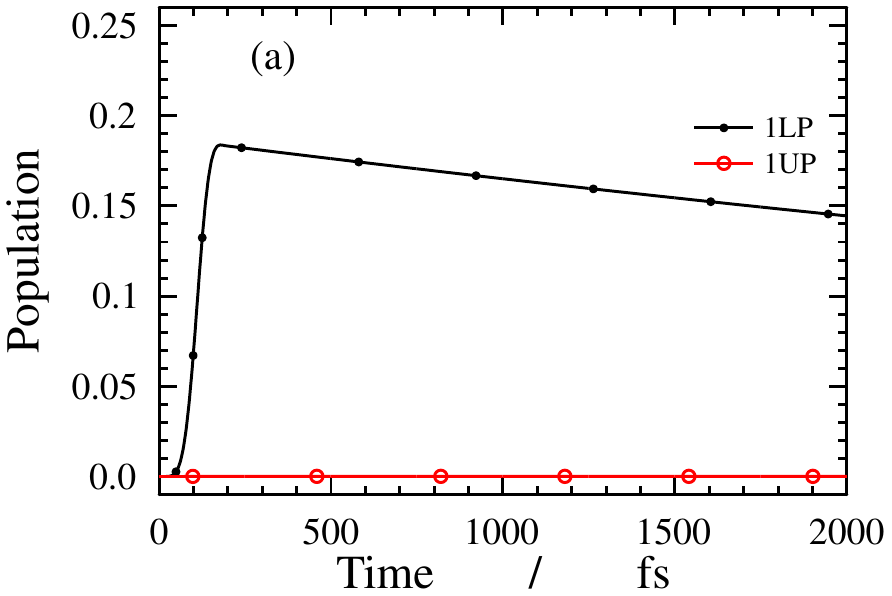}
\includegraphics[width=0.475\textwidth]{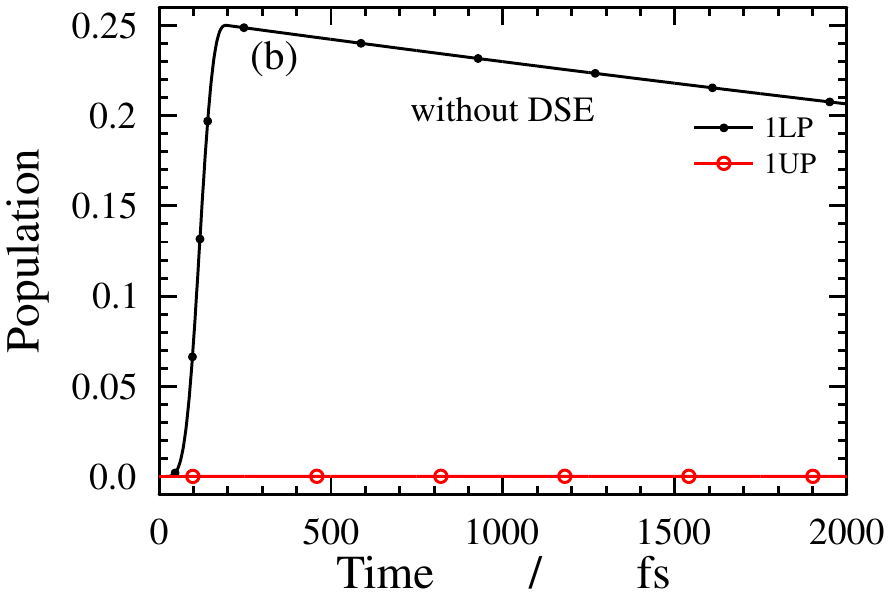}
\caption{\label{fig:ad_pop_2}
Populations of polaritonic states (1LP and 1UP) as a function of time obtained 
with (panel a) and without (panel b) the DSE term.
Cavity and laser parameters equal $\omega_\textrm{c}=34304.93~\textrm{cm}^{-1}$, $g=0.01~\textrm{au}$, $\gamma_\textrm{c}=10^{-4}~\textrm{au}$, 
$\mathbf{e} = (1,1,1)/\sqrt{3}$, $\omega=31389~\textrm{cm}^{-1}$, 
$I = 5~\textrm{TW/cm}^2$ and $T = 200~\textrm{fs}$.}
\end{figure}

\begin{figure}
\includegraphics[width=0.475\textwidth]{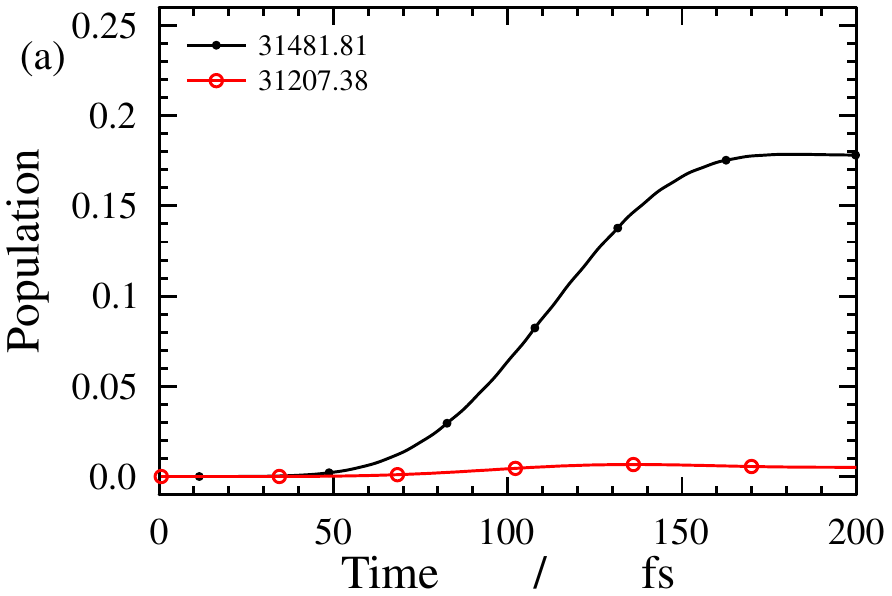}
\includegraphics[width=0.475\textwidth]{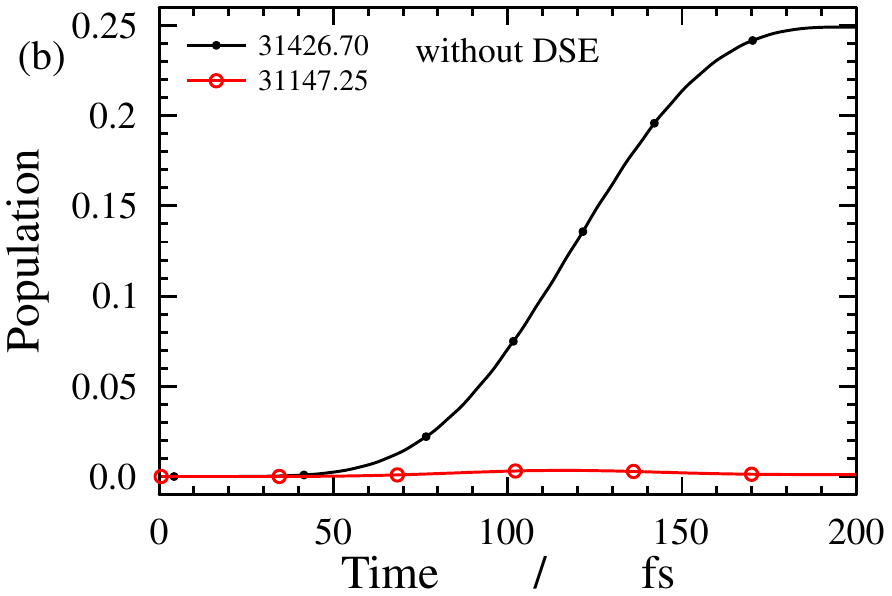}
\caption{\label{fig:ed_2}
Populations of cavity-molecule eigenstates populated by laser excitation
as a function of time (panel a: with DSE, panel b: without DSE).
Cavity and laser parameters equal $\omega_\textrm{c}=34304.93~\textrm{cm}^{-1}$, $g=0.01~\textrm{au}$, $\gamma_\textrm{c}=10^{-4}~\textrm{au}$, $\mathbf{e} = (1,1,1)/\sqrt{3}$, 
$\omega=31389~\textrm{cm}^{-1}$, $I = 5~\textrm{TW/cm}^2$ and $T = 200~\textrm{fs}$.
Energy levels in the plot legends are given in units of $\textrm{cm}^{-1}$ and 
referenced to the lowest energy level.}
\end{figure}

\begin{figure}
\includegraphics[width=0.475\textwidth]{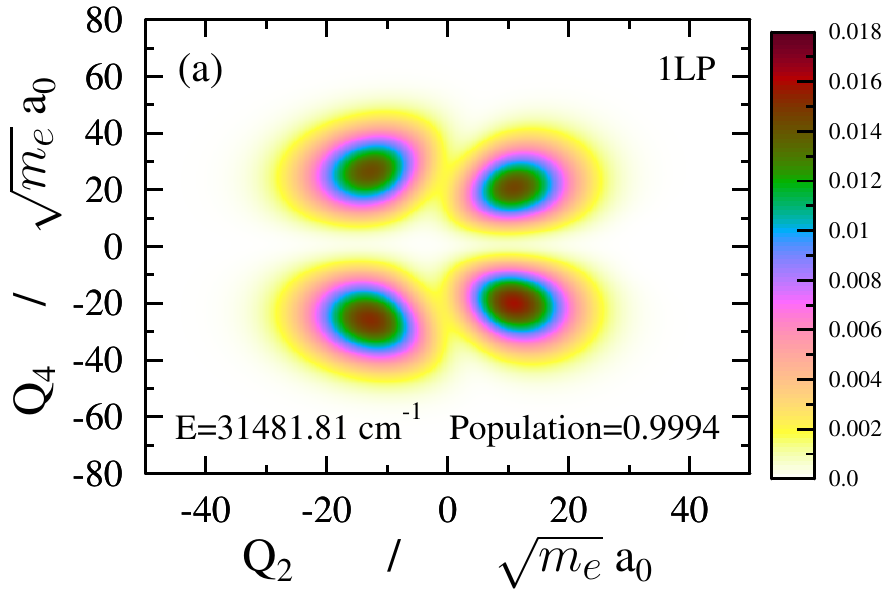}
\includegraphics[width=0.475\textwidth]{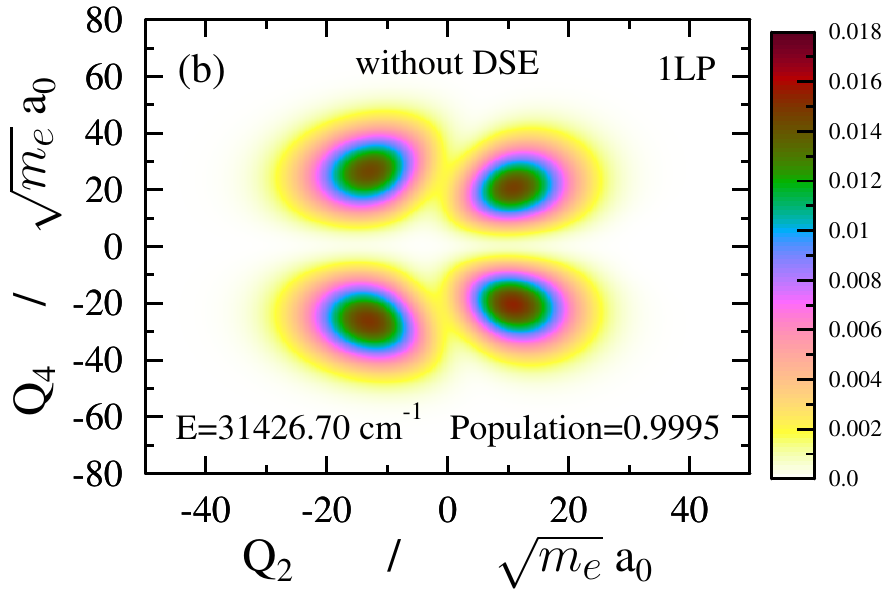}
\caption{\label{fig:dens_2}
Probability densities (1LP state) for cavity-molecule eigenstates 
populated by laser excitation (panel a: with DSE, panel b: without DSE).
Cavity parameters equal $\omega_\textrm{c}=34304.93~\textrm{cm}^{-1}$, $g=0.01~\textrm{au}$ 
and $\mathbf{e} = (1,1,1)/\sqrt{3}$.
$Q_2$ and $Q_4$ denote normal coordinates of the $\nu_2$ (C=O stretch) 
and $\nu_4$ (out-of-plane) normal modes.
Energy levels (given in the figures) are referenced to the lowest energy level.
Populations of the 1LP state are also given in each panel.
Note the similarity between panels a and b.}
\end{figure}

Next, the laser wavenumber is set to $\omega=31389~\textrm{cm}^{-1}$ while retaining
$\omega_\textrm{c}=34304.93~\textrm{cm}^{-1}$.
The corresponding results are presented in Figs. \ref{fig:ad_pop_2}, 
\ref{fig:ed_2} and \ref{fig:dens_2}, following the structure of results
shown previously. In this case, the laser pulse transfers population exclusively to
the 1LP polaritonic state. As shown in Fig. \ref{fig:ad_pop_2}, results obtained 
with or without DSE are qualitatively similar although the extent of population
transfer is affected. Fig. \ref{fig:ed_2}
reveals that dominantly one cavity-molecule eigenstate is populated by the laser
with ($31481.81~\textrm{cm}^{-1}$) and without ($31426.70~\textrm{cm}^{-1}$) DSE, 
in contrast to the previous resonant case. Moreover, the dominantly-populated eigenstates
can be well approximated by A vibrational eigenstates dressed with 0 photons, i.e.,
the laser excitation targets eigenstates which are not affected by resonant
cavity-molecule coupling. Fig. \ref{fig:dens_2} highlights the similarity between the
relevant 1LP probability densities of the eigenstates $31481.81~\textrm{cm}^{-1}$ and
$31426.70~\textrm{cm}^{-1}$ (both states have negligible 1UP populations).
 
\section{Summary and conclusions}
\label{sec:summary}

In this paper, we have presented a method for the computation of the dipole self-energy
(DSE) term. More precisely, we have implemented the computation of the matrix elements
of the squared dipole moment operator ($(\hat{\vec{\mu}}\vec{e})^2$ 
where $\vec{e}$ is the cavity polarization vector) in the basis of molecular 
electronic states of interest.
In this work, the ground and excited electronic states are
described using the RHF and CIS methods which have been found to give sensible results
for a two-dimensional vibrational model of the four-atomic formaldehyde molecule.
The extension to electronic states computed with more sophisticated approaches should
not pose any problem using the proposed method. 
Our method is similar to the one reported in Ref. \citenum{24FiRi} where 
squared dipole moment matrix elements were obtained for the electronic ground state at 
the RHF level of theory without any further approximations. 
In addition, our study complements earlier works based on
approximating the DSE term with the square of the ground-state dipole moment\cite{24YuBo}
and using the resolution-of-identity approach to treat the DSE.\cite{24BoScKo}
The latter might require the computation of several electronic states to
construct the approximate resolution of identity, while our procedure
necessitates only those electronic states which are coupled by the cavity mode
(electronic ground state and lowest singlet excited state in the current work).

With electronic structure data needed for the DSE at hand, we have investigated the 
effect of DSE on cavity-induced nonadiabatic quantum dynamics of a single molecule 
coupled to a quantized mode of an optical cavity.
We have found that inclusion of the DSE in the cavity-molecule Hamiltonian
changes both the spatial position and energy of the light-induced conical
intersection (LICI) between the 1LP (lower) and 1UP (upper) polaritonic potential 
energy surfaces in the so-called singly-excited subspace. 
It has also been shown that the DSE term  introduces a slight symmetry breaking 
which manifests in modifying the LICI position from $Q_4=0$ (without DSE) to 
$Q_4 \ne 0$ (with DSE) where $Q_4$ is the normal coordinate of the 
out-of-plane mode of formaldehyde.
We have presented time-dependent quantum-dynamical results which reflect the 
nonadiabaticity induced by the cavity. Without the cavity, formaldehyde does not 
possess conical intersections and its dynamics does not show nonadiabatic effects.
The molecule coupled to the cavity is pumped with a laser pulse which transfers
population from the lowest cavity-molecule eigenstate primarily to the 
singly-excited subspace.
Although the DSE leads to relatively small shifts in the energy levels and potentials,
LICIs are generally sensitive to even small changes of parameters and, consequently, 
the populations of the 1LP and 1UP polaritonic states as well as populations of 
cavity-molecule eigenstates can be rather sensitive to whether or not the DSE 
is taken into account. However, it is still possible to establish a one-to-one 
correspondence between cavity-molecule eigenstates obtained with or without the DSE,
as demonstrated for eigenstates populated by the laser excitation.
We note that small frequency shifts due to the DSE are also related 
to change in the refractive index.\cite{24FiRi}
The present investigation is for a single molecule and it would be of interest to 
extend it to study the impact of the DSE on the dynamics of a molecular ensemble 
in the cavity. 

\begin{acknowledgement}
The authors are indebted to NKFIH for funding (Grant No. K146096).
Financial support by the Deutsche Forschungsgemeinschaft (DFG) (Grant No. CE 10/56-1)
is gratefully acknowledged. The work performed in Budapest received funding from 
the HUN-REN Hungarian Research Network.
This paper was supported by the J\'anos Bolyai Research Scholarship
of the Hungarian Academy of Sciences. 
Supported by the University of Debrecen Program for Scientific Publication.
\end{acknowledgement}

\bibliography{cavity_dse}

\end{document}